\documentclass{jfm}
\usepackage{graphicx}
\usepackage{epstopdf, epsfig}
\usepackage{amsmath}
\usepackage{natbib}
\usepackage{amsbsy,amsfonts,amsgen,amssymb}
\usepackage{hyperref}
\usepackage{textcomp}
\usepackage{gensymb}
\usepackage{subcaption}
\usepackage{enumitem}

\title{
Hyperbolicity, shadowing directions and sensitivity analysis of a turbulent three-dimensional flow 
}
\author{
Angxiu Ni\aff{1} \corresp{\email{niangxiu@gmail.com}}
}
\affiliation{
\aff{1} Department of Mathematics, University of California, Berkeley, Berkeley, CA 94720, USA
}

\newcommand{\dd}[2]{\frac{d #1}{d #2}}

\newcommand{\avg}[1]{\left\langle #1 \right\rangle}

\newcommand{\R}{\mathbb{R}}

\newcommand{\mus}{{m_{us}}}
\newcommand{\integrate}{\int_{0}^{T}}

\begin{document}
\maketitle

\begin{abstract}
This paper uses compressible flow simulation to analyze the hyperbolicity, shadowing directions, and sensitivities of a weakly turbulent three dimensional cylinder flow at Reynolds number 525 and Mach number 0.1. 

By computing the first 40 Covariant Lyapunov Vectors (CLVs), we find that unstable CLVs are active in the near-wake region, whereas stable CLVs are active in the far-wake region.
This phenomenon is related to hyperbolicity since it shows that CLVs point to different directions; it also suggests that for open flows there is a large fraction of CLVs that are stable.
However, due to the extra neutral CLV and the occasional tangencies between CLVs, our system is not uniform hyperbolic.

By the Non-intrusive least-squares shadowing (NILSS) algorithm, we compute shadowing directions and sensitivities of long-time-averaged objectives.
Our results suggest that shadowing methods may be valid for general chaotic fluid problems.
\end{abstract}


\maketitle

\section{Introduction}

The chaotic dynamics of many fluid problems, such as turbulence and vortex streets, 
are central to many important challenges facing science and technology.
These chaotic flows are typically controlled by system parameters, such as the incoming flow conditions, or the boundary geometry.
This paper studies the three dimensional (3D) flow past a cylinder at Reynolds number 525 from a dynamical system point of view.
More specifically, we study perturbations in flow fields due to perturbations in initial conditions and system parameters,
which are covariant Lyapunov vectors (CLVs) and shadowing directions of this flow problem.
With the shadowing directions, we can then study perturbations in the long-time-averaged objectives due to perturbations in system parameters.

Chaotic fluid systems depend sensitively on initial conditions:
perturbations whose norms grow like exponential functions are called CLVs,
and their exponents are the Lyapunov Exponents (LEs) \citep{Katok_thick_book}.
Unstable CLVs determine the uncertainty in predicting the future behavior of the dynamical system,
provided the past trajectory is known up to some precision \citep{Young_entropy}.
More importantly, unstable CLVs can be regarded as chaotic degrees of freedom (DOFs) to which the apparent complicated behavior can be attributed.
This reduction in DOFs allows reduced-order numerical methods, such as those developed by \citet{Chorin_ROM,Duraisamy_ROM}.
In experiments, we can infer some properties of CLVs by reconstructed attractors \citep{Sieber_CLV_experiments},
but more details can be obtained only numerically.
In particular, \citet{Moin_LE_PoiseuilleFlow} computed the LE spectrum of turbulent Poiseuille flow.
Recently, the LE spectrum for some 3D computational fluid dynamics (CFD) problems have been studied \citep{Blonigan2016b,P.Fernandez2017}.

There remain questions about active areas of CLVs in `open' flows, or developing flows such as boundary layers, jets and wakes.
Since perturbations in open flows are typically convected downstream, it is difficult to form an intuitive visualization of the CLVs.
In contrast, for `closed' flows such as Benard convection and Taylor-Couette flow, global dynamics are expressed at all spatial locations,
and CLVs can be effectively interpreted as eigenmodes \citep{Moin_LE_PoiseuilleFlow}.
\citet{Fernandez_LE_discretization} plotted the flow fields of CLVs in an open flow, 
and they found no significant difference in active areas among CLVs, perhaps because only a few CLVs were computed.
However, in this paper, we are able to observe the difference in active areas for different CLVs.

Hyperbolicity is a property of dynamical systems which says that the unstable and stable CLVs are bounded away from each other.
Uniform hyperbolicity is assumed in many theoretical results, such as 
the existence of a steady distribution \citep{young2002srb}, linear response formula \citep{Ruelle_diff_maps,Ruelle_diff_flow},
and shadowing lemma \citep{Bowen_shadowing,Pilyugin1999book}.
However, many chaotic fluid systems are not uniform hyperbolic.
For a Kolmogorov flow simulated with 224 DOFs \citep{Inubushi2012}
and for a 3D Boussinesq equations simulated with $5\times 10^5$ DOFs \citep{Xu2016},
researchers found for both cases that the angles between CLVs can be very small.
Currently, there are not many results regarding hyperbolicity for CFD simulated 3D Navier-Stokes fluid systems.
In this paper, we conclude that our fluid system is not uniform hyperbolic; 
however, we find a more robust phenomenon related to hyperbolicity.

It has been conjectured that many theoretical results, although mathematically developed under uniform hyperbolicity, 
are still valid in most chaotic systems
\citep{gallavotti_chaotic_hypothesis_1995, gallavotti_chaotic_hypothesis_2006,ruelle_turbulence_hypothesis}.
Support for such conjectures can be found in \citep{Albers2006,Reick2002,Ni_NILSS_JCP}.
In this paper, we are particularly interested in whether shadowing methods are still valid, even though our system is not uniform hyperbolic.

For a chaotic system, if we coordinate carefully perturbations on both initial conditions and parameters, 
we can find a shadowing trajectory which remains close to the old trajectory \citep{Bowen_shadowing}.
The first-order approximation of the difference between the two trajectories is the shadowing direction \citep{Pilyugin_shadowing_diffeomorphism}.
Shadowing methods enable sensitivity analysis of long-time-averaged objectives \citep{Wang_ODE_LSS},
which is useful in design \citep{Jameson1988}, control \citep{Bewley2001}, inverse problems \citep{Tromp}, 
error estimation \citep{Becker2001,Giles2002,Fidkowski}, data assimilation\citep{Thepaut1991},
and training neural networks \citep{deeplearning_book_Goodfellow,linearRange_GD}.
Shadowing is difficult to observe experimentally, and the only method we are aware of is numerical.
The theory for shadowing methods typically assumes uniform hyperbolicity;
however, in this paper, we show that shadowing directions also exist for our fluid system.

The least-squares shadowing (LSS) method \citep{wang2014convergence, Wang_ODE_LSS} 
computes sensitivities of long-time-averaged objectives via computing the shadowing direction.
With high cost, LSS has been successfully applied in two-dimensional (2D) CFD problems \citep{Blonigan2016}.
The non-intrusive least-squares shadowing (NILSS) method \citep{Ni_NILSS_AIAA_2016, Ni_NILSS_JCP} reformulates LSS
by constraining computation to the unstable subspace.
For real-life problems, such as 2D flow over a step computed in \citep{Ni_NILSS_JCP}, 
NILSS can be thousands times faster than LSS \cite{Blonigan2016}.
One variant of NILSS is the adjoint, such as one developed by \citet{Blonigan_2017_adjoint_NILSS}, 
and the non-intrusive least-squares adjoint shadowing (NILSAS) algorithm developed by \citet{Ni_nilsas}.
In particular, NILSAS is based on new theoretical progress in adjoint shadowing directions \citep{Ni_adjoint_shadowing}.
Another variant is the Finite Difference NILSS (FD-NILSS) algorithm \citep{Ni_fdNILSS}, whose implementation requires only primal solvers.
NILSS and its variants have been successfully used for several other turbulent fluid system, 
such as a 2D flow over a backward step \citep{Ni_NILSS_JCP},
and a minimal flow unit of a channel flow \citep{Blonigan_2017_adjoint_NILSS};
NILSS has also been successfully applied to easier mathematical models such as the Lorenz 63 and Kuramoto–Sivashinsky (KS) systems.
In this paper, we use FD-NILSS to compute shadowing directions and sensitivities of our problem.

We begin the main part of the paper by describing the physical and numerical set-up for our flow problem.
Then the rest of the paper consists of two logically connected parts.
First, we review LEs and CLVs, and compute LEs and CLVs for our flow problem.
Although the CLV result violates the uniform hyperbolicity, 
it still indicates some form of hyperbolicity, which encourages us to proceed to the second part of the paper,
where we review shadowing directions and the FD-NILSS algorithm,
using which we compute shadowing directions and sensitivities of several long-time-averaged objectives.

\section{Problem set-up and verification of simulation}

\begin{figure}\begin{center}
  \includegraphics[width = 0.6\textwidth]{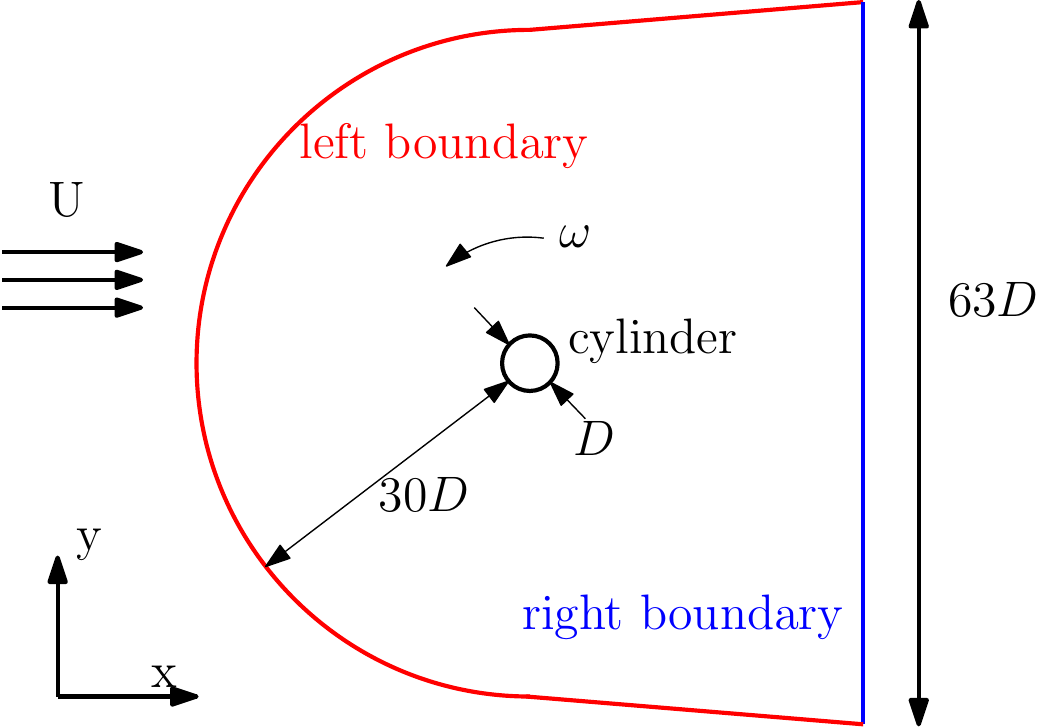}
  \caption{Geometry used in the simulation of a 3D flow past a cylinder. 
    The spanwise extent of the computational domain is $Z=2D$. 
    The positive direction of the cylinder rotational speed $\omega$ is counter-clockwise.}
  \label{f:geometry}
\end{center}\end{figure}

Our physical problem of the 3D flow past a cylinder is the same as in \citep{Ni_fdNILSS}.
The front view of the geometry of the entire flow field is shown in figure \ref{f:geometry}.
In the rest of the paper, the units we have in mind are SI units, but since in this paper all values are normalized,
and in the software all equations are dimensionless, readers may take any compatible set of units.
The diameter of the cylinder is $D=0.25\times10^{-3}$.
The spanwise width is $Z=2D$.
The free-stream conditions are: density $\rho_0=1.18$, pressure $P_0 = 1.01\times 10^5$, 
temperature $T_0=298$, dynamic viscosity $\mu=1.86\times10^{-5}$.
The free-stream flow is in the $x$-direction, with the velocity $U$ being one of the system parameters, and for the base case $U_0=33.0$.
The flow-through time $t_0$, defined as the time for $U_0$ flowing past the cylinder, is $t_0 = D/U_0 = 7.576\times 10^{-6}$.
The Reynolds number of the base case is $Re=\rho_0 U_0 D /\mu = 525$ and the Mach number is $0.1$.
The cylinder can rotate around its center with rotational speed $\omega$, which is the second system parameter for our problem.
The $\omega$ value is measured in revolutions per unit time, and its positive direction is anticlockwise, as shown in figure~\ref{f:geometry}.
For the cylinder to rotate one cycle per flow-through time, the rotation speed $\omega_0=1/t_0=1.32\times 10^5$.

\begin{figure}
  \centering
  \includegraphics[trim=10cm 2cm 12cm 1cm, clip=true, width=0.49\textwidth]{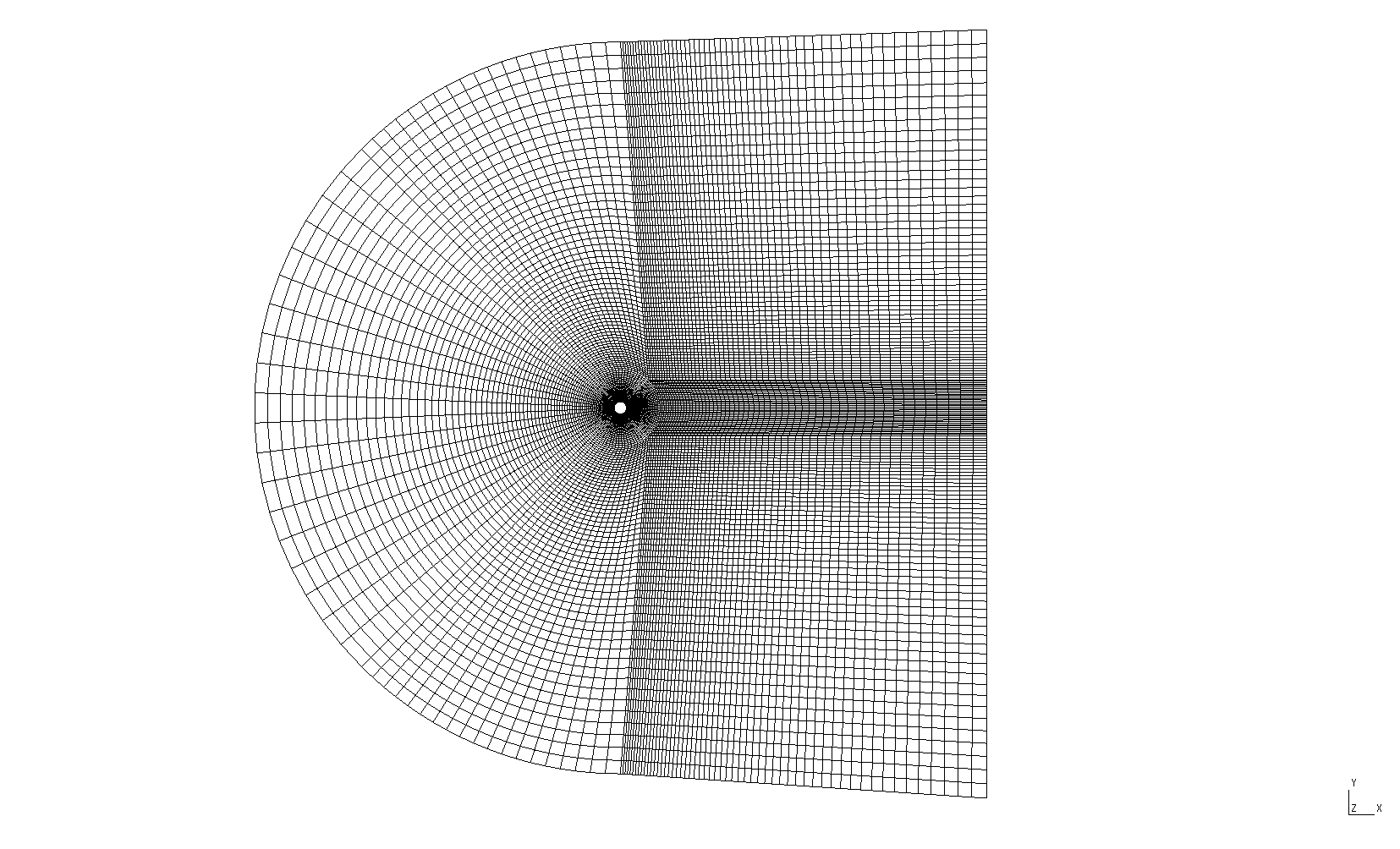}
  \includegraphics[trim=10cm 0cm 10cm 0cm, clip=true, width=0.49\textwidth]{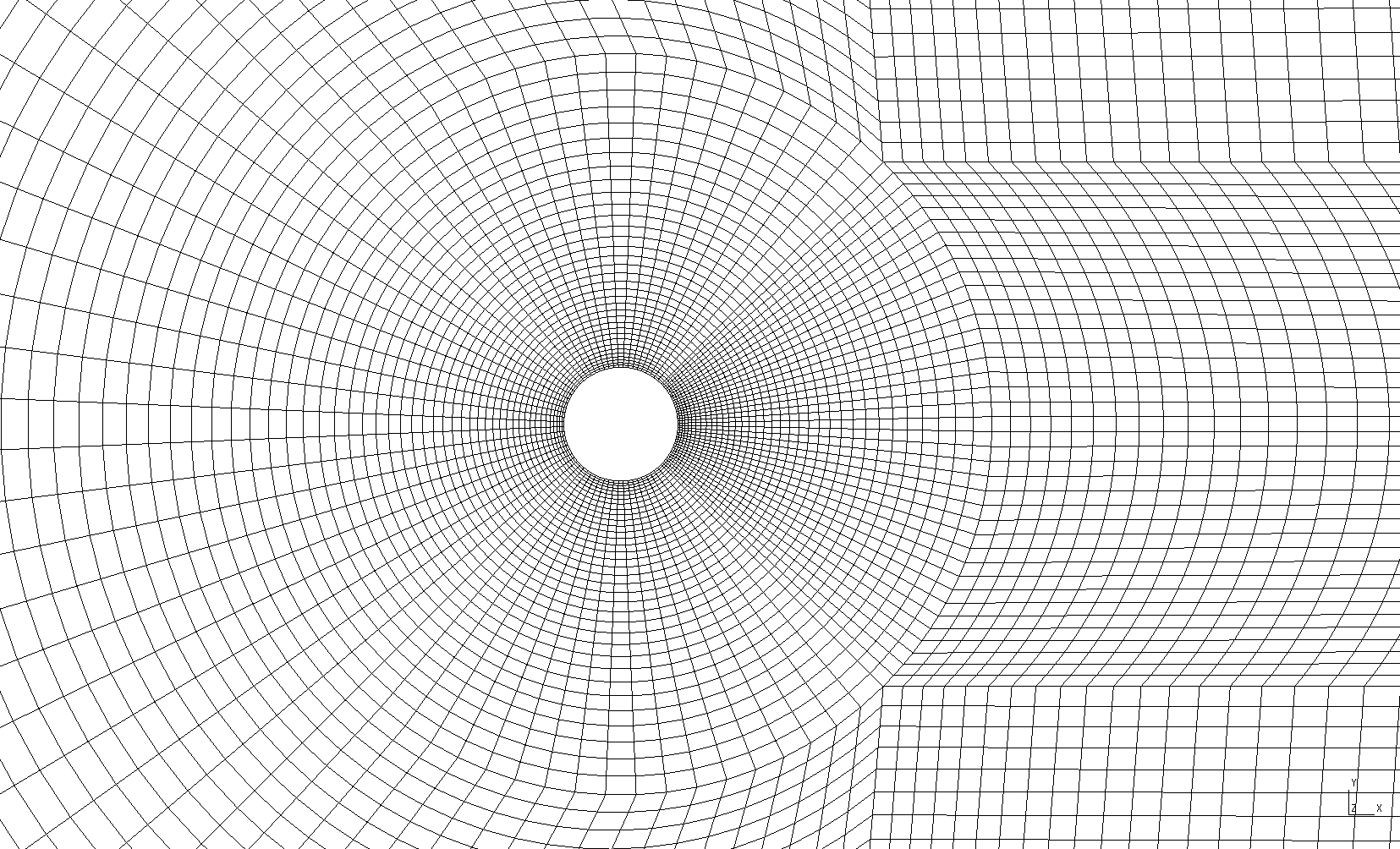}
  \caption{Left: 2D slice of the finer mesh over the entire computational domain. 
    Right: zoom around the cylinder. 
    This is a block-structured mesh with $7.5\times10^5$ hexahedra.
    The spanwise direction has 48 cells.}
  \label{f:mesh}
\end{figure}

We run our simulations on two sets of block-structured hexahedral meshes.
The coarser mesh has $3.7\times10^5$ cells, and finer mesh has $7.5\times10^5$ cells.
A 2D slices of the finer mesh is shown in figure \ref{f:mesh}.
For both meshes, the spanwise direction has 48 cells.
The CFD solver we use is CharLES developed at Cascade Technologies \citep{bres_Charles_solver}, 
using which we perform under-resolved direct numerical simulation (uDNS) without turbulence model.
The accuracy of the solver is formally second order in space and third order in time.
The spanwise boundaries use periodic boundary conditions.
The left boundary uses a convective boundary condition \citep{Colonius1993};
the right boundary uses the Navier-Stokes characteristic boundary conditions (NSCBC) \citep{Poinsot1992}.
The time-step size is $\Delta t = 10^{-8} = 1.32\times 10^{-3} t_0$.

\begin{figure}
  \centering
  \includegraphics[trim=0cm 8cm 0cm 8cm, clip=true, width = 0.85\textwidth]{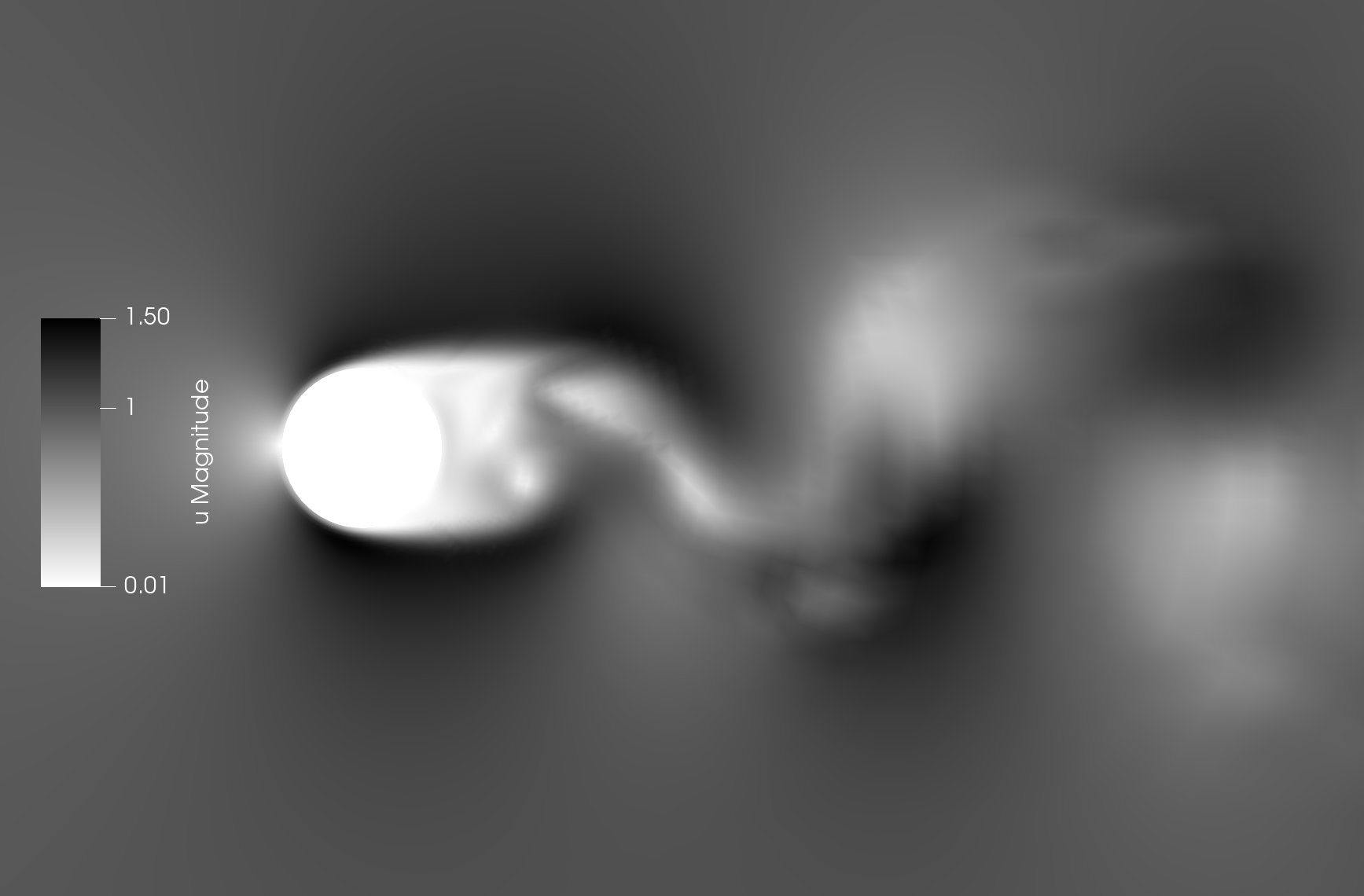}\\
  \includegraphics[trim=0cm 12cm 0cm 12cm, clip=true, width = 0.85\textwidth]{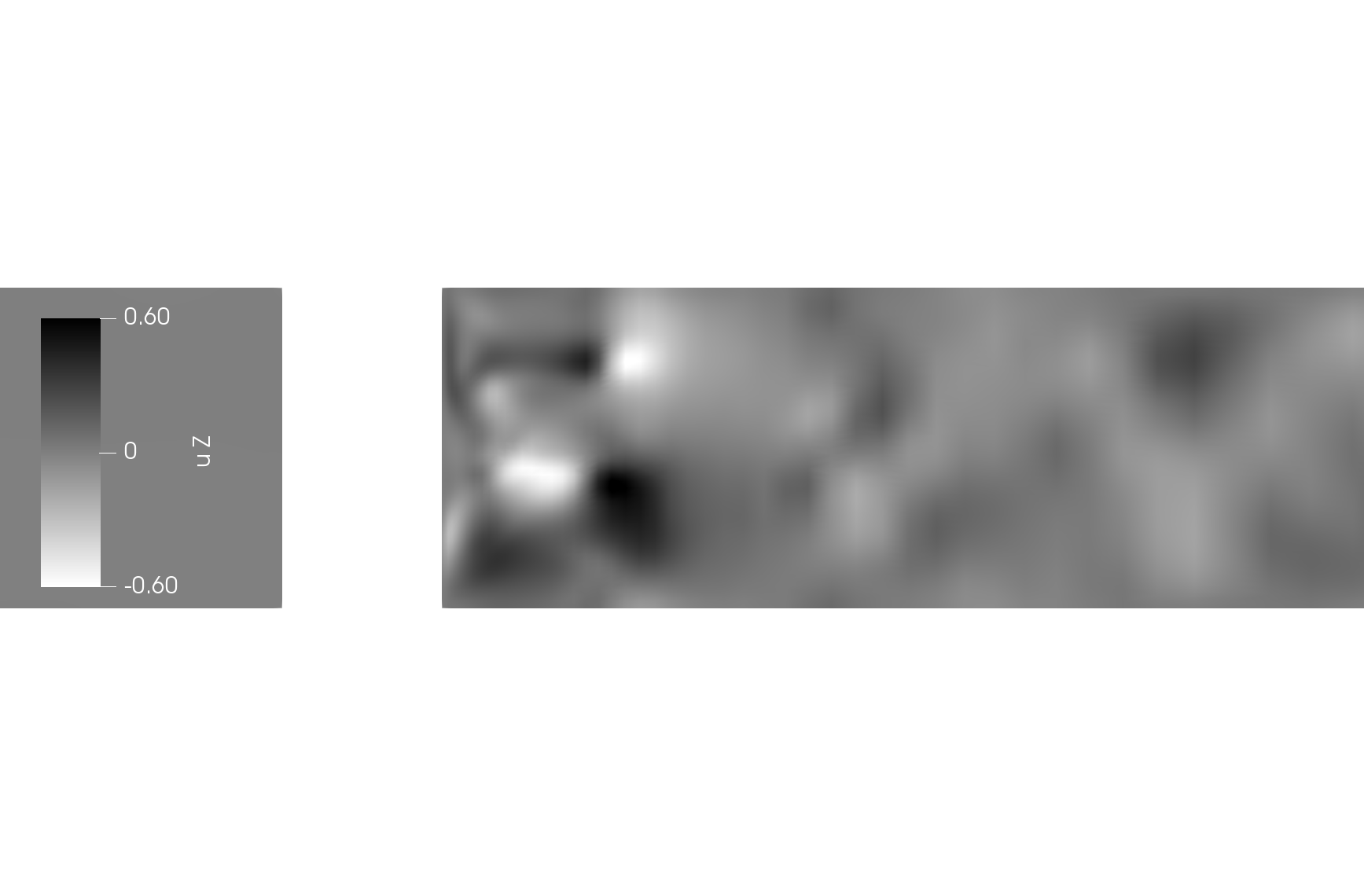}
  \caption{2D slices of the flow field on the finer mesh. 
    Top: vertical cross-section, plotted by the magnitude of velocity.
    Bottom: horizontal cross-section, plotted by the spanwise velocity.
    The bottom picture shows the flow is 3D.
    All velocities are normalized by the free-stream velocity $U_0=33.0$.}
  \label{f:flow field}
\end{figure}

\begin{figure}
  \centering
  \includegraphics[trim=0cm 12cm 0cm 12cm, clip=true, width = 0.85\textwidth]{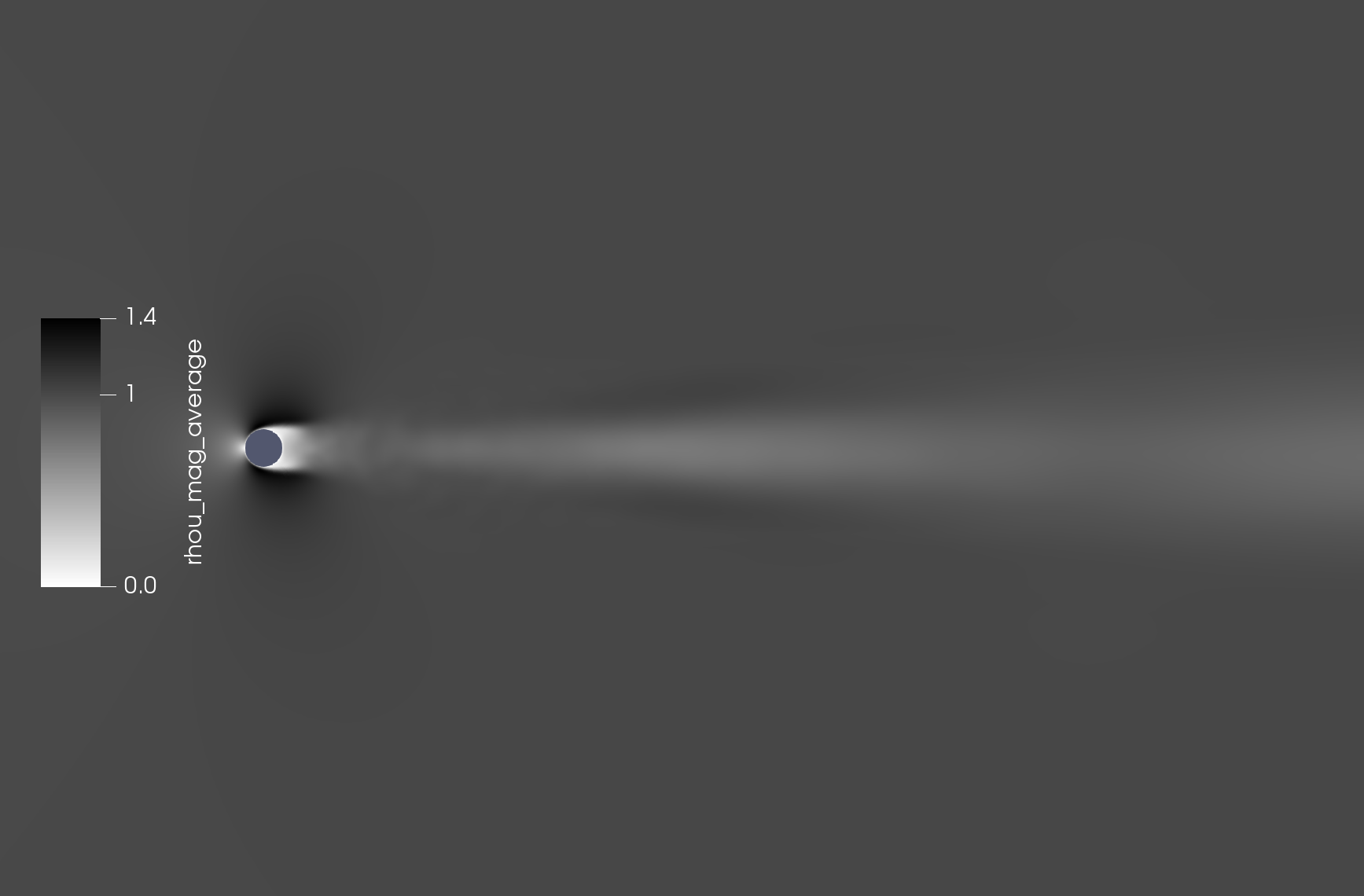}\\
  \caption{Flow field on the finer mesh averaged over a time span of 168$t_0$,
    plotted by the $\rho U$ component, normalized by $\rho_0 U_0$.}
  \label{f:primal averaged}
\end{figure}

Some 2D snapshots of the flow field simulated with the above numerical settings on the finer mesh are shown in figure \ref{f:flow field}.
The flow is chaotic and 3D.
We also plotted a time-averaged flow field in figure~\ref{f:primal averaged}.
As we can see, the computed flow field generally fits our fluid intuitions in the stagnant area before and after the cylinder,
the acceleration area around the cylinder, and the boundary layer and wake structures.

We compare our results with previous literature.
The same physical problem has been investigated in experiments by \citet{williamson1990measurements}, 
and in numerical simulations by \citet{Mittal1996}.
We compare the Strouhal number $S_t$ and the averaged drag coefficient $C_D$.
Here the Strouhal number is defined by $S_t=fD/U$, where $f$ is the main frequency of the vortex shedding, represented by the history of the lift;
the drag coefficient $C_D = D_r / (0.5\rho_0 U^2 D Z)$, where $D_r$ is the drag.
As shown in table \ref{t:verify}, our results on both meshes match previous experimental and numerical results.

\begin{table}
  \setlength{\tabcolsep}{12pt}
  \centering
  \begin{tabular}{r l l }
    & $S_t$ & $C_D$ \\
    Coarser mesh with $3.7\times10^5$ cells                   & 0.21 	& 1.22  \\
    Finer mesh with 	 $7.5\times10^5$ cells                & 0.21 	& 1.19  \\ 
    2D simulation by \citet{Mittal1996} 		      & 0.22 	& 1.44 	\\
    3D simulation by \citet{Mittal1996}		      & 0.22 	& 1.24 	\\
    experiment by \citet{williamson1990measurements}    & 0.21 	& 1.15 	\\
  \end{tabular}
  \caption{Comparison with previous literature in the Strouhal number $S_t$ and the averaged drag coefficient $C_D$.}
  \label{t:verify}
\end{table}

\section{Covariant Lyapunov vectors and hyperbolicity}

\subsection{Definitions of primal and tangent solutions} \label{s:define tangents}

Both CLVs and shadowing directions are tangent solutions,
which describe evolutions of trajectory perturbations caused by perturbations on a dynamical system. 
There are two kinds of perturbations we can perform on a dynamical system given in equation~(\ref{e:dynamical_system}): 
perturbation on initial conditions and on system parameters.
The corresponding tangent solutions are homogeneous and inhomogeneous tangent solutions, respectively.
As we shall see, CLVs are special homogeneous tangent solutions, while shadowing directions are special inhomogeneous tangent solutions.

We write the governing equation of the flow field in the form of a general dynamical system, which will be referred to as the primal system:
\begin{equation} \label{e:dynamical_system}
  \dd{u}{t} = f(u,s), \quad u \rvert_{t=0} = u^0 + v^{0} s + w^0\phi \,.
\end{equation}
Here $f(u,s):X\times \R\rightarrow X$ is a smooth function, $u$ are state variables, and $s$ is the system parameter.
The initial condition is $u^0$ with a possible perturbation in the direction of $v^{0} $ or $ w^0$, controlled by $s$ and another parameter $\phi$.
A primal solution $u:\R\rightarrow X$ maps time to a compact phase space $X$, 
where a single point represents an entire 3D flow field at an instant;
thus $u(t)$ can as well be imagined as a trajectory in $X$.

We assume the system is autonomous, that is, $f$ does not depend on time.
We could extend our discussions to some special non-autonomous cases,
such as those with some compactness in the time dependence, so that the ergodic theory is still available.
For example, for cases whose time dependence can be decomposed into finitely many frequencies,
we can extend the phase space by adding finitely many tori and still maintain a compact phase space.
On this extended phase space our analysis will still be valid;
for example, \cite{antali2018periodicLSS} used the shadowing method to compute sensitivities.
However, our discussions in this paper do not extend to non-autonomous systems in the most general setting.

Our dynamical system considered in this paper is given by the semi-discretized Navier-Stokes equation without turbulence models.
More specifically, on the spatial direction, the flow field, together with boundary conditions, 
is represented by the finite-volume method on our meshes.
The state variables are conservative variables, that is, $\rho$, $\rho U$, and $\rho E$.
Hence our phase space $X \subset \R^m$ is finite-dimensional, with $m$ being the number of DOFs.
We further assume that $X$ is a bounded hence compact subset.
For the coarser mesh, the number of DOFs is approximately $1.9\times 10^6$; 
for the finer mesh, it is approximately $3.8\times 10^6$.
On the other hand, we regard the temporal direction as being continuous;
thus our dynamical system is continuous, and later the theories we use are for continuous dynamics.

We can differentiate equation~(\ref{e:dynamical_system}) with respect to $\phi$, and define $w= du / d\phi $.
Then $w$ satisfies the so-called homogeneous tangent equation:
\begin{equation} \label{e:homo tangent}
  \dd{w}{t} - \partial_u f w = 0 \,, \quad w(t=0) = w^0\,,
\end{equation}
where $\partial_u f \in \R^{m\times m}$, and the ordinary differential equation (ODE) is called the homogeneous tangent ODE.
Here $w$ reflects the trajectory perturbation caused by perturbing the initial condition in the direction of $w^0$.
On the other hand, we can differentiate \eqref{e:dynamical_system} with respect to $s$, and define $v= du/ds$.
Then $v$ satisfies the inhomogeneous tangent equation: 
\begin{equation} \label{e:inhomo tangent}
  \dd{v}{t} - \partial_u f v = \partial_s f \,, \quad v(t=0) = v^0\,,
\end{equation}
where $\partial_s f \in \R^m$, and the ODE is called the inhomogeneous tangent ODE.
Here $v$ reflects the trajectory perturbation caused by perturbing the system parameter $s$, 
which not only affects the governing differential equation,
but also potentially impacts the initial condition in the direction of $v^0$.
The solution set of an inhomogeneous tangent ODE can be written as a particular inhomogeneous $v^*$ plus the solution set of the homogeneous tangent ODE,
which is a linear subspace due to homogeneity.

\subsection{Definitions of covariant Lyapunov vectors and hyperbolicity} \label{s:define CLVs and hyperbolicity}

A CLV $\zeta(t)$ is a homogeneous tangent solution whose norm behaves like an exponential function of time. 
That is, there are $C_1,C_2>0$ and $\lambda\in \R$ such that, for any $t$,
\begin{equation}\label{eq:CLV}
  C_1 e^{\lambda t }\|\zeta(u(0))\|  \le \|\zeta(u(t))\| \le C_2 e^{\lambda t }\|\zeta_j(u(0))\| ,
\end{equation}
where the norms are Euclidean norms in $\R^m$, and $\lambda$ is defined as the LE corresponding to this CLV. 
CLVs with positive LEs are said to be unstable, CLVs with negative LEs are stable, and CLVs with zero LEs are neutral.
The existence of CLVs is given by the Oseledets theorem \citep{oseledets1968multiplicative}, for which a proof in English can be found in \citep[chapter 4]{Rezakhanlou2018},
and more reference can be found at \citep{Oseledets_scholarpedia}.
In this paper, we sort LEs by decreasing order, so the $j$-th largest LE and its corresponding CLV will be called the $j$-th LE and $j$-th CLV, respectively.

Hyperbolicity is roughly defined as being that the tangent space at every state splits into a stable subspace, 
an unstable space, and a neutral subspace.
The one most widely used in ergodic theory is uniform hyperbolicity, which has two assumptions: 
1) the three subspaces are uniformly bounded away from each other;
and 2) the neutral subspace is one-dimensional.
Our system cannot be uniform hyperbolic, since it violates the first assumption, that is, it has at least two neutral CLVs:
the first one corresponds to the common time translation of continuous dynamical systems,
and the second corresponds to spanwise translations due to the periodic boundary conditions.

Still, we want to check if our system violates the second assumption in uniform hyperbolicity.
More specifically, we want to bound the norm of the operator which projects to one of the subspaces,
i.e. the $C_\alpha$ given in \citep{Ni_adjoint_shadowing}, 
where we showed that $C_\alpha = (1-\cos^2 \alpha)^{-1/2}$, with $\alpha$ being the smallest angle between subspaces.
Of course, the ideal case would be that these subspaces are orthogonal to each other,
leading to $\alpha=90^\circ$ and $C_\alpha=1$ takes its minimum.
Both the bound of the norm of the shadowing direction \cite{Pilyugin_shadowing_diffeomorphism}
and the bound of the error in FD-NILSS \citep{Ni_NILSS_JCP} are proportional to $C_\alpha$, 
and we do not want it larger than 10, causing impact of an order of magnitude.
The corresponding angle is $\alpha_0=5.7^\circ$, which serves as our threshold.

As we will see, the smallest angle between all CLVs can be affected by mesh or time length.
However, we find a robust phenomenon related to hyperbolicity, that is, 
angles between two CLVs become much larger if their indices are further apart.
We no longer have uniform hyperbolicity, since there is more than one neutral CLV and there may be tangencies between adjacent CLVs.
On the other hand, this phenomenon is more robust, and reveals more structures of CLVs not indicated even by uniform hyperbolicity.
In fact, we can even visualize this phenomenon by showing that active areas of CLVs are different.

More importantly, we will show that for our problem, where we do not have uniform hyperbolicity, tools given by ergodic theories are still valid,
although they mostly use uniform hyperbolicity as logical assumptions.
Specifically, we can still compute shadowing directions which correctly reflect sensitivities of statistics of the fluid system.
Our results confirm the chaotic hypothesis \citep{Gallavotti:2008}, 
which says that many tools logically consequential to the uniform hyperbolicity assumption are in fact valid for a larger class of chaotic systems.

\subsection{Numerical methods for computing Lyapunov exponents and covariant Lyapunov vectors} \label{s:algorithm_CLV}

The algorithm we use to compute LEs is given by \citet{Benettin1980_LE}, 
and the algorithm for CLVs is by \citet{Ginelli2007_CLV,Ginelli2013_CLV}.
Since these two algorithms share many of the same procedures, we describe them together.
For $i=0,\dots, K-1$, we define the $i$-th time segment as time span $[t_i, t_{i+1}]$, with $t_i=i\Delta T$.
In the algorithm presented below, for quantities defined on the entire segments such as $u_i$ and $W_i$, 
we use the same subscript as the segment they are defined on.
For quantities defined only at interfaces between segments such as $Q_{i}$ and $R_{i}$, we use the same subscript as the time point they are defined at.

To begin, we should prescribe: 
(1) number of LEs/CLVs to compute, $M$; 
(2) length of each time segment, $\Delta T$;
and (3) number of time segments, $K$.
Consequently, the time length of the entire trajectory, $T=K\Delta T$, is determined.
Then the algorithm is given by the following procedure.

\begin{enumerate}
  \item Generate initial conditions for primal solutions and homogeneous tangent solutions.

  \begin{enumerate}
    \item Compute the primal solution of equation (\ref{e:dynamical_system}) 
      for sufficiently long time so that the trajectory lands onto the attractor, 
      then set $t=0$, and set initial condition of the primal system, $u_0(0)$.

    \item Randomly generate an $m\times M$ orthogonal matrix $Q_{0}= [ q_{01}, \dots, q_{0M}]$.
      This $Q_0$ will be used as initial condition for homogeneous tangent solutions.
  \end{enumerate}

  \item For $i=0$ to $K-1$, on segment $i$, where $t\in[t_i,t_{i+1}]$, do:

  \begin{enumerate}
    \item Compute the primal solution $u_i(t)$ from $t_i$ to $t_{i+1}$.

    \item Compute homogeneous tangent solutions $W_i(t) = [ w_{i1}(t), \dots, w_{iM}(t) ]$.

    \begin{enumerate}
      \item For each homogeneous tangent solution $w_{ij}$, $j=1, \dots, M$, starting from initial condition $w_{ij}(t_i) = q_{ij}$,
        integrate equation~\eqref{e:homo tangent} from $t_i$ to $t_{i+1}$.

      \item Perform QR factorization: 
        $W_i(t_{i+1}) = Q_{i+1} R_{i+1}$, where $Q_{i+1} = [q_{i+1,1}, \dots, q_{i+1,M}]$.
    \end{enumerate}
  \end{enumerate}

  \item The $j$-th largest LE, $\lambda_j$,  is approximated by
    \begin{equation}
      \lambda_j = \frac{1}{K\Delta T}\sum_{i=1}^{K} \log |D_{ij}| \,,
    \end{equation}
    where $D_{ij}$ is the $j$-th diagonal element in $R_i$.
    This formula for $\lambda_j$ converges to the true value as $T$ becomes large.

  \item We define an $m\times M$ matrix $V(t)$ by
    \begin{equation}
      V(t) = W_i(t) R^{-1}_{i+1} \dotsb R^{-1}_{K} \,,\, t\in [t_i, t_{i+1}] \,.
    \end{equation}
    The $j$-th column of $V(t)$ converges to the direction of the $j$-th CLV when both $t$ and $(T-t)$ become large.
    Notice that, although $V(t)$ has different expressions on different segments, its columns are continuous across all segments.
\end{enumerate}

We discuss the inner product we implicitly used in defining the orthogonality of the $Q$ matrices.
The inner product between two flow fields at an instant is a summation of inner products between different components.
For consistency of units, we normalize each component by free-stream properties.
Moreover, we should prescribe a geometric metric for inner products between the same components.
This metric should have bounded total volume, so that inner products are finite;
it should also put more weight closer to the cylinder, so that FD-NILSS has better accuracy for the surface objectives we use later.
A natural selection of such a metric is simply one where all mesh cells are equally weighted, 
and the entire space outside the mesh has weight zero.
More specifically,
with subscript $l$ denoting different components, and $k$ denoting different cells of the mesh,
for $v, w\in X=\R^m$,
we define inner products as
\begin{equation} \label{e:define_inner_products}
  v^T w := \frac 1 {m'} \sum_{k=1}^{m'} v_{k}^T w_{k} \,, 
  \textnormal{ where} \quad
  v_{k}^T w_{k} := \frac 13 \left(\frac{v_{k1} w_{k1}}{\rho_0^2} 
  + \frac{\sum_{l=2}^4 v_{kl} w_{kl}}{\rho_0^2U_0^2}
  + \frac{v_{k5} w_{k5}}{\rho_0^2E_0^2} \right)\,.
\end{equation}
Here $m'$ is the number of cells.
We order the 5 components of states by $\rho$, $\rho U$, $\rho E$, and normalize all components of $\rho U$ by $\rho U_0$.
Under our definition, if every point in a flow field has the same state as the free-stream, then this flow field has unit norm.

The convergence of CLV in the above algorithm is in terms of the distance between one-dimensional subspaces.
Since a CLV is a homogeneous tangent solution, we can multiply it by any factor, 
and still get a homogeneous tangent solution whose norm behaves like an exponential function with the same LE, 
which, by definition, is still the same CLV.
In other words, it is only the directions of CLVs that are meaningful, but not the magnitudes.
Hence we can normalize CLVs by any factor we like.
In this paper, we normalize CLVs such that their maximal values are 1.0.

In this paper, we use finite differences to approximate tangent solutions. 
To approximate a homogeneous solution $w$ with initial condition $w^0$, 
we use the definition $w = du/d\phi\approx \Delta u/ \Delta \phi$,
where $\phi$ controls the perturbation in the initial condition in the direction of $w^0$.
More specifically, we compute a perturbed primal solution $u^w$ with a perturbed initial condition $u^0 + \Delta \phi$, 
and the approximation for $w$ is $w \approx (u^w- u)/\Delta \phi$, where $u$ is the base trajectory.
\footnote{The python package `fds' implementing FD-NILSS is available on GitHub via \href{https://github.com/qiqi/fds}{this link}.
  It can also perform Lyapunov analysis.
  The particular files related to the application in this section are in fds/apps/charles\_cylinder3D\_Lyapunov.}

\subsection{Results of Lyapunov exponents} \label{s:LE results}

The time-step size used for computing LEs and CLVs is the same as in the numerical simulation,
that is, $\Delta t = 10^{-8} = 1.32\times 10^{-3} t_0$.
The number of homogeneous tangent solutions we compute is $M=40$.
Each segment has 200 time steps, which gives $\Delta T = 2\times 10^{-6} = 0.264t_0$.
The total number of time segments is $K=800$, hence the time length of the entire trajectory is $T = 1.6\times 10^{-3} = 211t_0$.

\begin{figure}
  \centering
  \centering
  \begin{subfigure}{\textwidth}
    \includegraphics[trim=0cm 0cm 0cm 0cm, clip=true, width=0.49\textwidth]{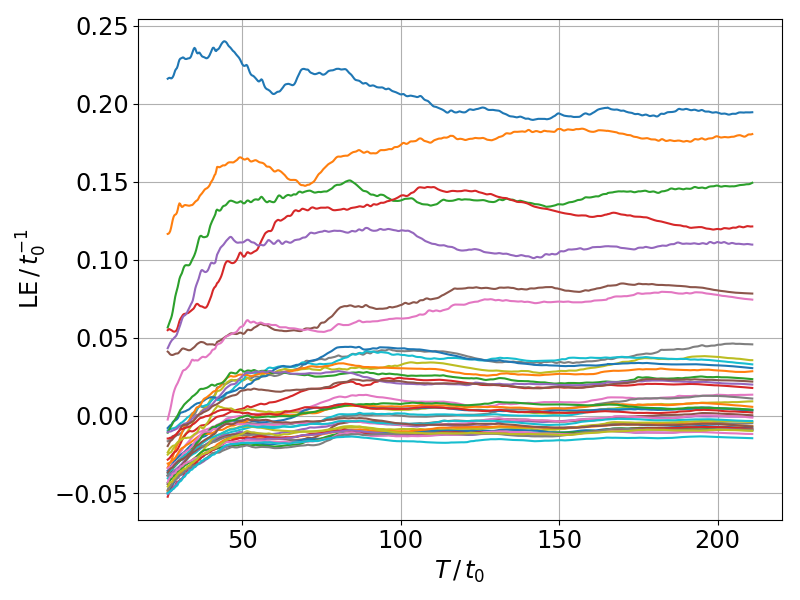}
    \includegraphics[trim=0cm 0cm 0cm 0cm, clip=true, width=0.49\textwidth]{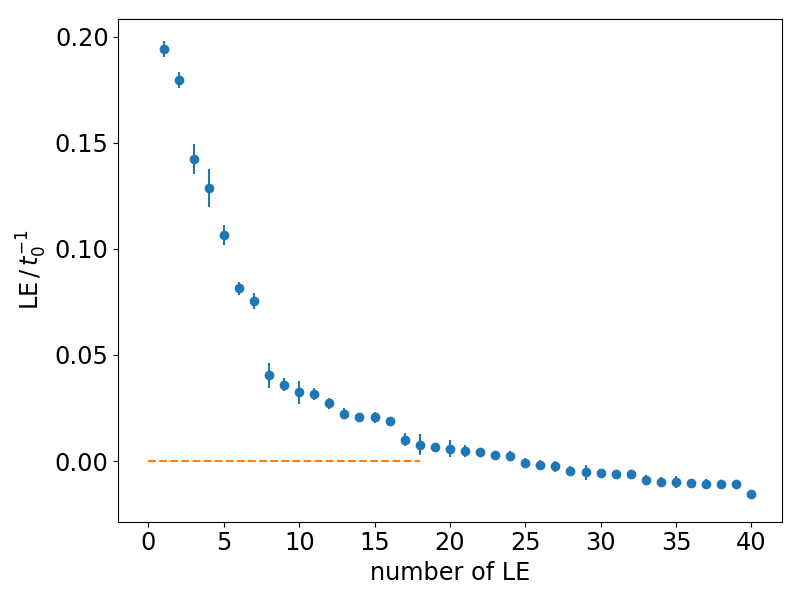}
    \caption{Results on the coarser mesh. The largest LE is 0.20$t_0^{-1}$, 
    meaning in one flow-through time $t_0$, the norm of the first CLV becomes $e^{0.20}=1.22$ times larger.}
  \end{subfigure}
  \begin{subfigure}{\textwidth}
    \includegraphics[trim=0cm 0cm 0cm 0cm, clip=true, width=0.49\textwidth]{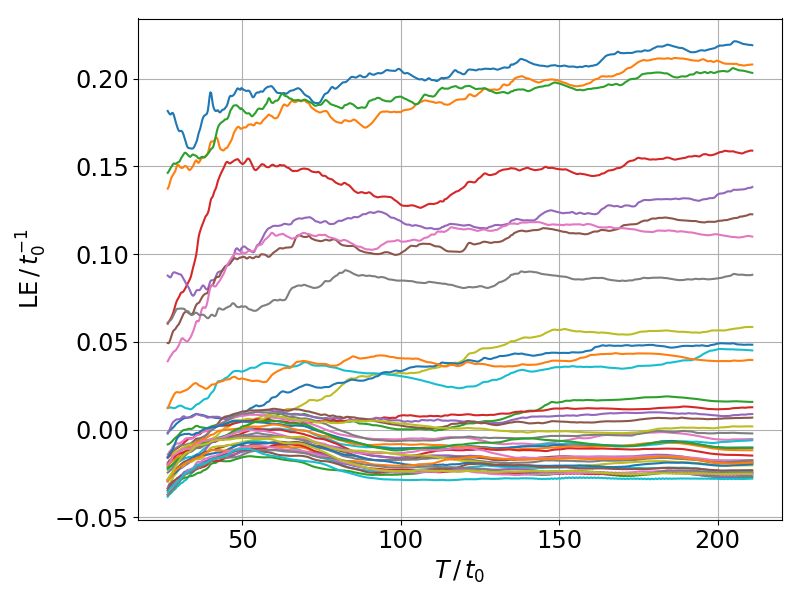}
    \includegraphics[trim=0cm 0cm 0cm 0cm, clip=true, width=0.49\textwidth]{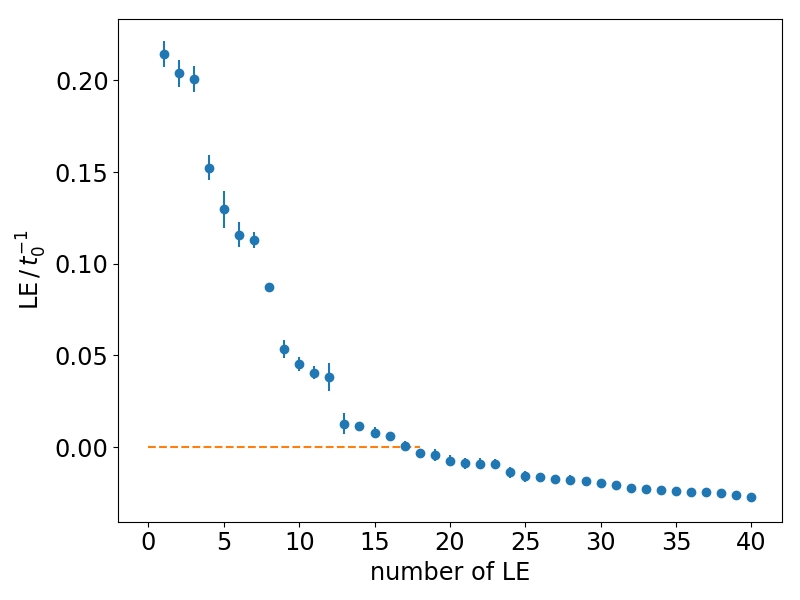}
    \caption{Results on the finer mesh. The largest LE is 0.21$t_0^{-1}$.}
  \end{subfigure}
  \caption{Lyapunov exponents (LE) normalized by $t_0^{-1}$. 
    Left: convergence history of the first 40 different LEs.
    Right: confidence intervals of LEs.} 
  \label{f:LE}
\end{figure}

With the above numerical settings, the convergence history of the first 40 LEs is shown on the left of figure~\ref{f:LE}.
As explained in \citep{Ni_NILSS_JCP}, 
the confidence interval of an LE is estimated by the smallest interval which bounds the history of the LE and whose size shrinks as $T^{-0.5}$.
The first 40 LEs and their confidence intervals are shown on the right of figure~\ref{f:LE}.

We can see that LE results depend on the meshes, as similarly discussed by \citet{Fernandez_LE_discretization}.
In figure~\ref{f:LE}, although the general shape of the LE spectrum looks similar,
a closer look will show that the finer mesh has larger but fewer positive LEs.
As a result, the selection of neutral CLVs is ambiguous and changes due to meshes.

However, we will show that there are several important physical phenomena which, 
although they are logically consequential to the Lyapunov analysis results,
exist robustly regardless of the detailed value of the LE spectrum.
This agrees with the chaotic hypothesis \citep{Gallavotti:2008}, 
which says that many tools logically consequential to the uniform hyperbolicity assumption
are in fact valid for a larger class of dynamical systems.
More specifically, those phenomena are as follows:
\begin{enumerate}
  \item the dimension of the attractor is much lower than the dimension of the system;
  \item angles between far-apart CLVs are large, and, in particular, CLVs are active at different locations;
  \item shadowing directions exist, and reflect the sensitivities of statistics to parameter perturbations.
\end{enumerate}
The focus of this paper will be the last two phenomena.
Moreover, for engineering interests, we will show that FD-NILSS can compute shadowing directions and, 
further, the sensitivities with reasonable cost.

We first show that the dimension of the attractor is much lower than the dimension of the system.
Intuitively, to remain its invariant hyper-volume, 
the attractor manifold should contain enough contracting (stable) CLVs to balance out the expansion caused by unstable CLVs.
More specifically, \citep{Kaplan_Yorke_dimension} estimated the attractor dimension by the so-called Lyapunov dimension:
\begin{equation} \label{e:K_Y_dimension}
  D_\lambda = N + \frac{1}{|\lambda_{N+1}|}\sum_{j=1}^{j=N}\lambda_j \in [N,N+1] \,,
\end{equation}
where $N$ is such that $\lambda_1+\lambda_2+\dots+\lambda_N>0$ and $\lambda_1+\lambda_2+\dots+\lambda_{N+1}< 0$.
We want to give a bound to $N$ on the finer mesh.
To do this, we remind readers that the algorithm we are using computes LEs by descending order, 
so we have $\lambda_j\le\lambda_{40}$ for all $j\ge40$.
Using the first condition of $N$, we have
\begin{equation}
  \sum_{j=1}^{40}\lambda_j + (N-40)\lambda_{40} = 
  \sum_{j=1}^{40}\lambda_j + \sum_{j=41}^{N}\lambda_{40}
  \ge \sum_{j=1}^{40}\lambda_j + \sum_{j=41}^{N}\lambda_{j} 
  >0 \,,
\end{equation}
which further leads to
\begin{equation}
  N \le 40 + \frac{\sum_{j=1}^{40}\lambda_j}{|\lambda_{40}|} = 40 +  \frac{1.031}{0.027} = 78 \,.
\end{equation}
On the other hand, since $\lambda_1+\cdots \lambda_{40} >0$, by the second property of $N$, we know $N> 40$.
Hence the Lyapunov dimension $41 \le D_\lambda\le 79$.
In other words, the chaotic dynamic of our flow problem on the finer mesh can be attributed to the interaction of fewer than 79 degrees of freedom.
Using the same method, we can estimate that the attractor dimension on the coarser mesh is smaller than 109.

\subsection{Results of covariant Lyapunov vectors and hyperbolicity} \label{s:CLV results}

\begin{figure}
  \centering
  \begin{subfigure}{0.47\textwidth}
    \includegraphics[trim=0cm 0cm 0cm 0cm, clip=true, width=\textwidth]{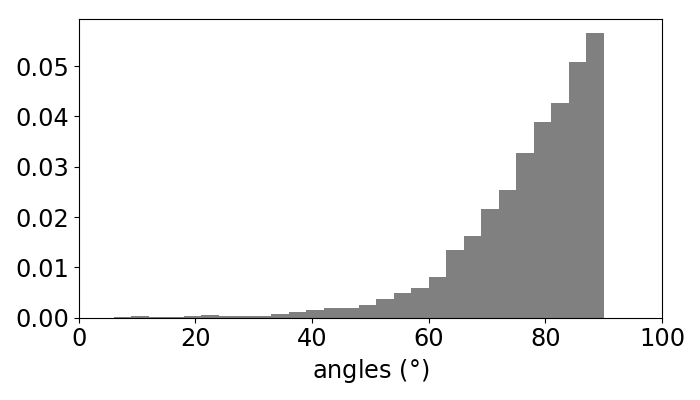}
    \caption{Coarser mesh, all CLVs, smallest angle is 5.8\degree.}
  \end{subfigure}
  \hfill
  \begin{subfigure}{0.47\textwidth}
    \includegraphics[trim=0cm 0cm 0cm 0cm, clip=true, width=\textwidth]{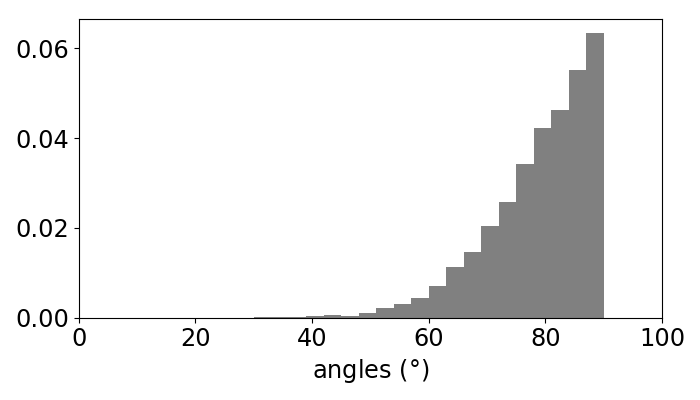}
    \caption{Coarser mesh, CLVs indices more than 5 apart, smallest angle is 29.2\degree}
  \end{subfigure}
  \begin{subfigure}{0.47\textwidth}
    \includegraphics[trim=0cm 0cm 0cm 0cm, clip=true, width=\textwidth]{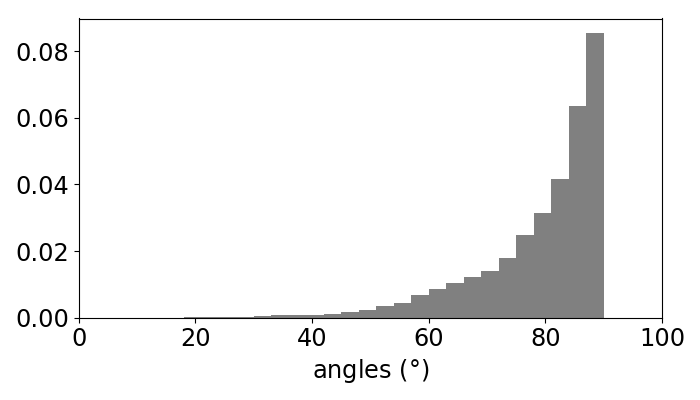}
    \caption{Finer mesh, all CLVs, smallest angle is 16.4\degree.}
  \end{subfigure}
  \hfill
  \begin{subfigure}{0.47\textwidth}
    \includegraphics[trim=0cm 0cm 0cm 0cm, clip=true, width=\textwidth]{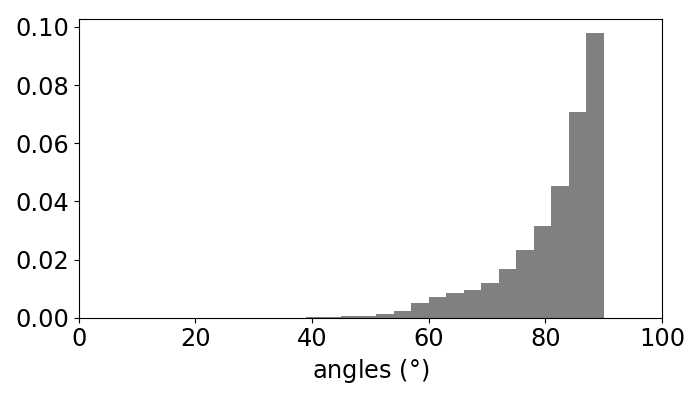}
    \caption{Finer mesh, CLVs indices more than 5 apart, smallest angle is 40.8\degree.}
  \end{subfigure}
  \caption{Histogram of angles between CLVs on segments 350 to 449,
    with the entire trajectory of length $T = 211t_0$ partitioned into $K=800$ segments. 
    The total area of the histogram is normalized to 1.}
  \label{f:angles all}
\end{figure}

In the left of figure~\ref{f:angles all}, we plot the histogram of angles between all pairs of the first 40 CLVs,
which are perturbations whose norm grows exponentially, as defined in equation~\eqref{eq:CLV}.
Notice the range of angles is $[0\degree, 90\degree]$, 
since the angle between the directions of two CLVs is the angle between subspaces of dimension one.
For the finer mesh, the smallest angle is 16.4\degree.
However, for the coarser mesh, the smallest angle is only 5.8\degree, almost equal the threshold value of 5.7\degree.
This indicates that the smallest angle and hence the second assumption in uniform hyperbolicity
is not a robust criterion to check for fluid systems, 
since it is plausible that the smallest angle falls below the threshold for another mesh or a longer time span.

Above dependencies on meshes call for a closer examination to find some robust physical phenomena not depending on meshes.
A deeper look into CLV results finds that tangencies happen only among adjacent CLVs: 
this result is similar to that obtained by \citet{Xu2016}.
Hence we plot the angles between CLVs whose indices are more than five apart.
For example, for the 20th CLV, we consider its angles with the 1st to 15th CLVs and 25th to 40th CLVs.
This result is plotted in the right of figure~\ref{f:angles all}.

As we can see, when the exponents of CLVs are further apart, their angles become larger: 
this phenomenon is related to hyperbolicity but not formally named yet.
So, on one hand, what we observe violates the second assumption of uniform hyperbolicity since occasional tangencies may fail the threshold.
But, on the other hand, our observation is even stronger than the uniform hyperbolicity,
because it specifies a trend not suggested by the definition of uniform hyperbolicity.

\begin{figure}
  \centering
  \begin{subfigure}{0.47\textwidth}
    \includegraphics[trim=0cm 0cm 0cm 0cm, clip=true, width=\textwidth]{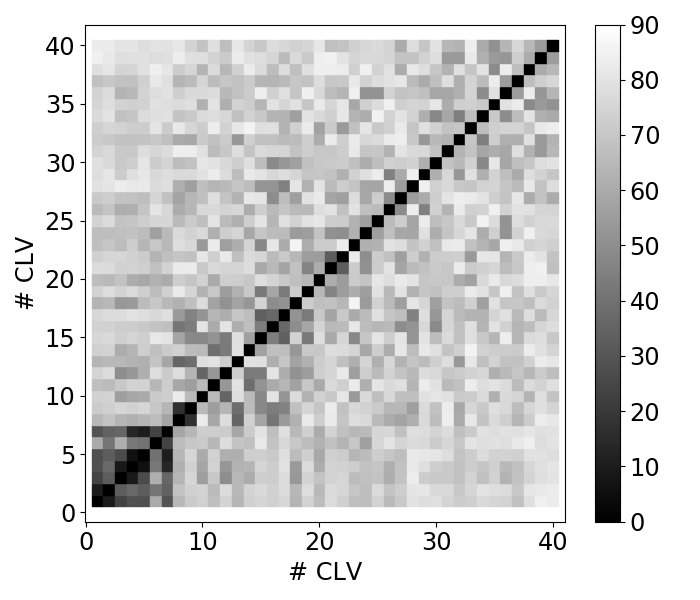}
    \caption{Coarser mesh, minimal angles between all pairs of CLVs.}
  \end{subfigure}
  \hfill
  \begin{subfigure}{0.47\textwidth}
    \includegraphics[trim=0cm 0cm 0cm 0cm, clip=true, width=\textwidth]{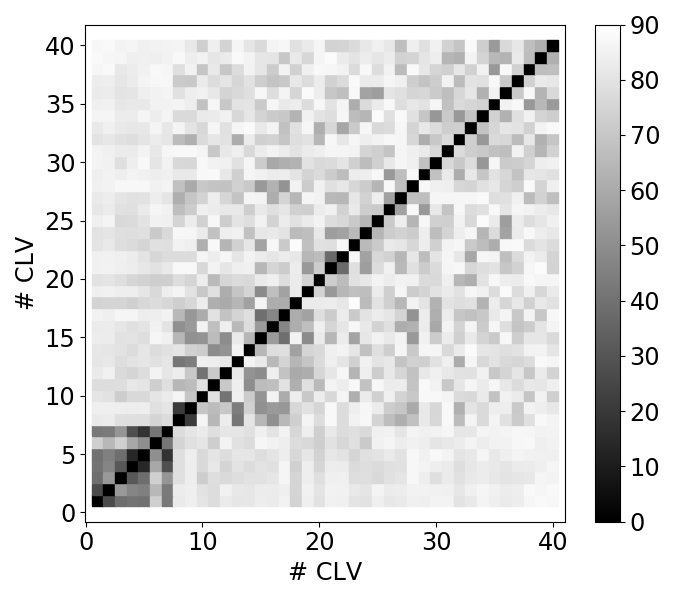}
    \caption{Coarser mesh, averaged angles between all pairs of CLVs.}
  \end{subfigure}
  \begin{subfigure}{0.47\textwidth}
    \includegraphics[trim=0cm 0cm 0cm 0cm, clip=true, width=\textwidth]{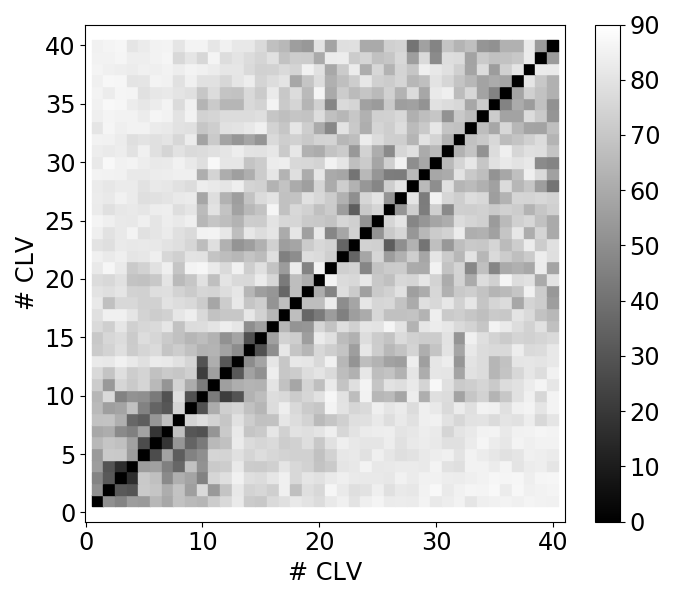}
    \caption{Finer mesh, minimal angles between all pairs of CLVs.}
  \end{subfigure}
  \hfill
  \begin{subfigure}{0.47\textwidth}
    \includegraphics[trim=0cm 0cm 0cm 0cm, clip=true, width=\textwidth]{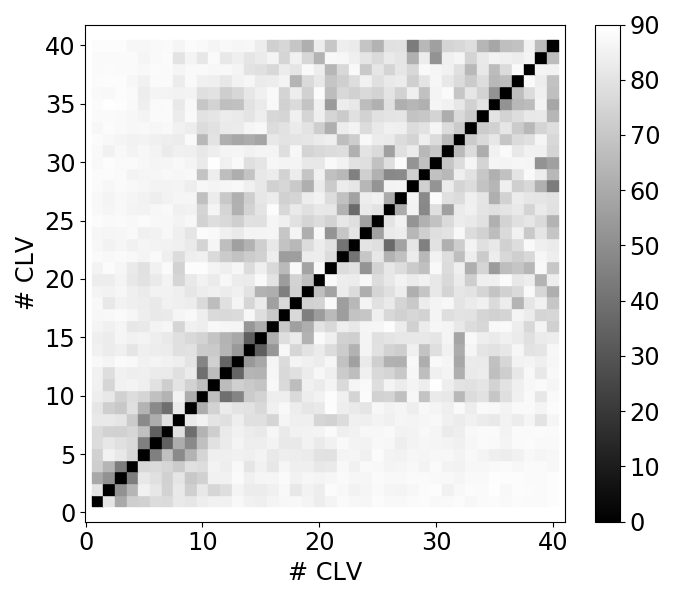}
    \caption{Finer mesh, averaged angles between all pairs of CLVs.}
  \end{subfigure}
  \caption{Minimal and averaged angles between all pairs of CLVs.
    The minimum and average are taken on segments 350 to 449, with the entire trajectory of length $T = 211t_0$ partitioned into $K=800$ segments.}
  \label{f:angles square}
\end{figure}

To further illustrate, we plot minimal and averaged angles between all pairs of CLVs in figure~\ref{f:angles square}.
Not surprisingly, most tangencies happen between adjacent CLVs, and the angles between CLVs get larger for CLVs whose indices are farther away.
In particular, on the coarser mesh, the first seven CLVs are closely coupled, 
whereas the first several CLVs on the finer mesh are less tangent to each other.
In general, tangencies among CLVs are rare events for both meshes, but even less frequent for the finer mesh.
Once again, the rareness of the tangencies indicates that smallest angle and hence uniform hyperbolicity is not a robust criterion;
on the other hand, the fact that angles between far-apart CLVs are large is robust to meshes,
and this may lead to some definition of a new form of hyperbolicity.

\begin{figure}
  \centering
  \begin{subfigure}{\textwidth}
    \includegraphics[trim=4cm 13cm 8cm 13cm, clip=true, width=0.49\textwidth]{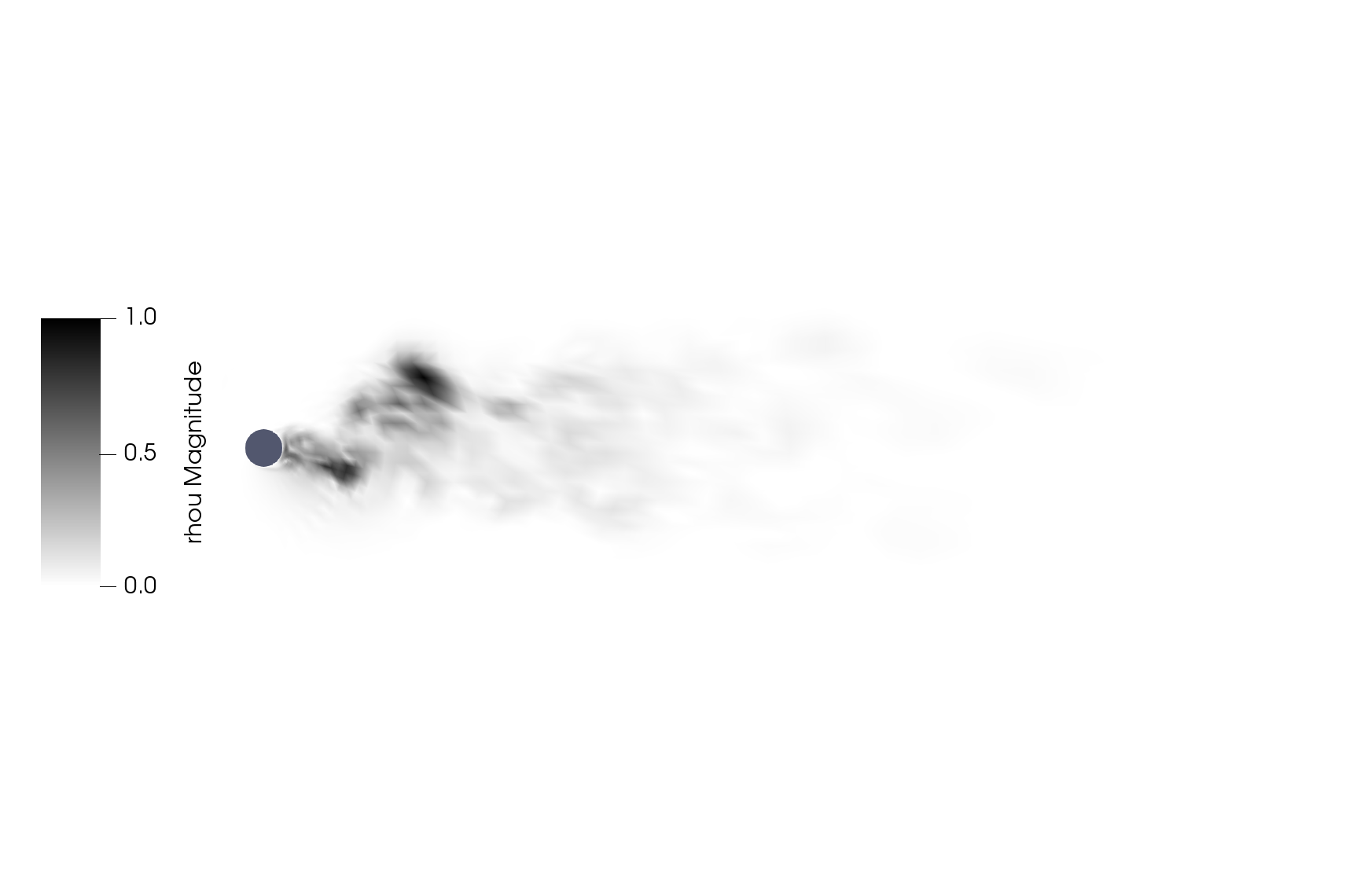}
    \includegraphics[trim=4cm 13cm 8cm 13cm, clip=true, width=0.49\textwidth]{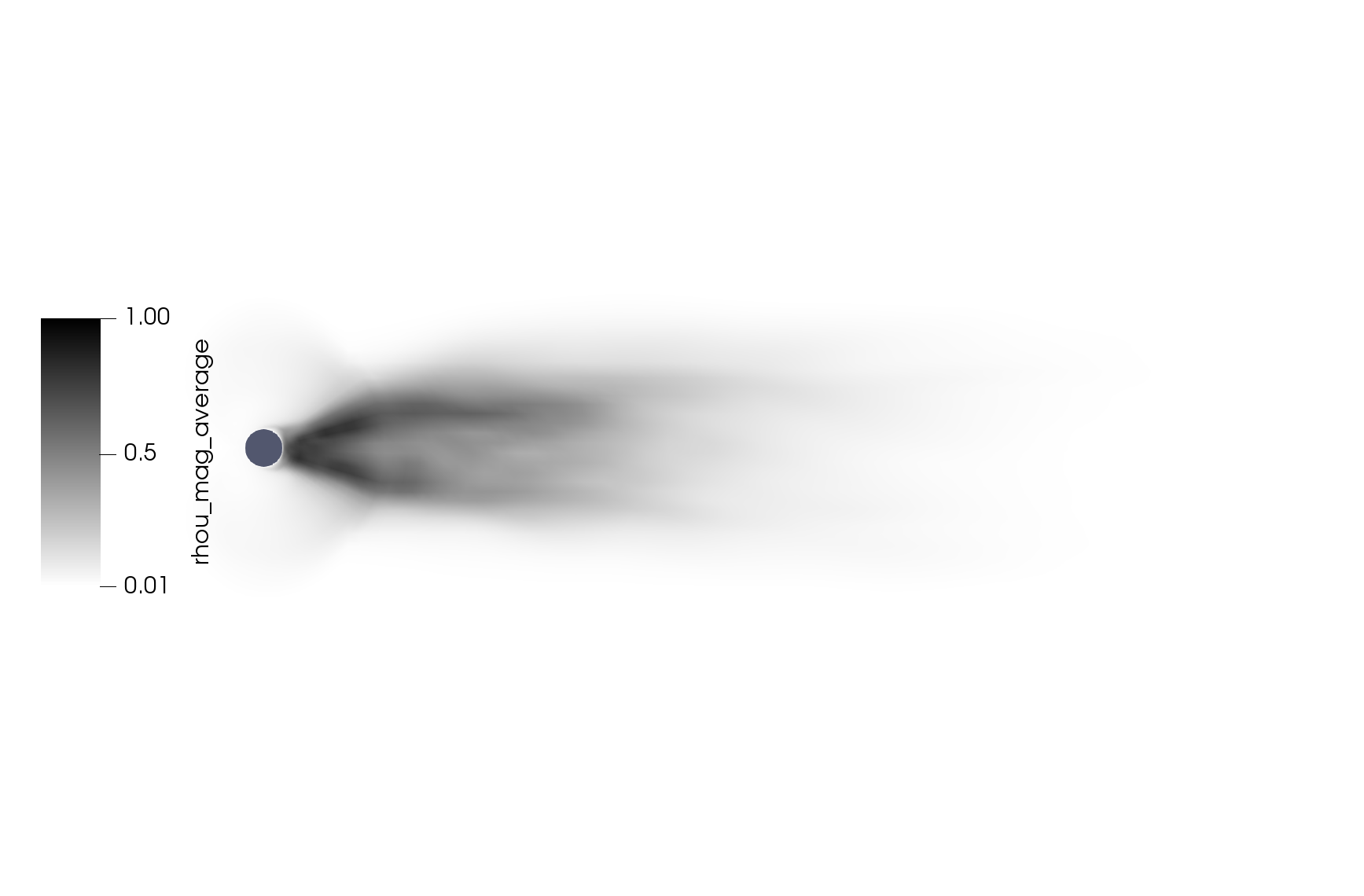}
    \caption{1st CLV}
  \end{subfigure}
  
  \begin{subfigure}{\textwidth}
    \includegraphics[trim=4cm 13cm 8cm 13cm, clip=true, width=0.49\textwidth]{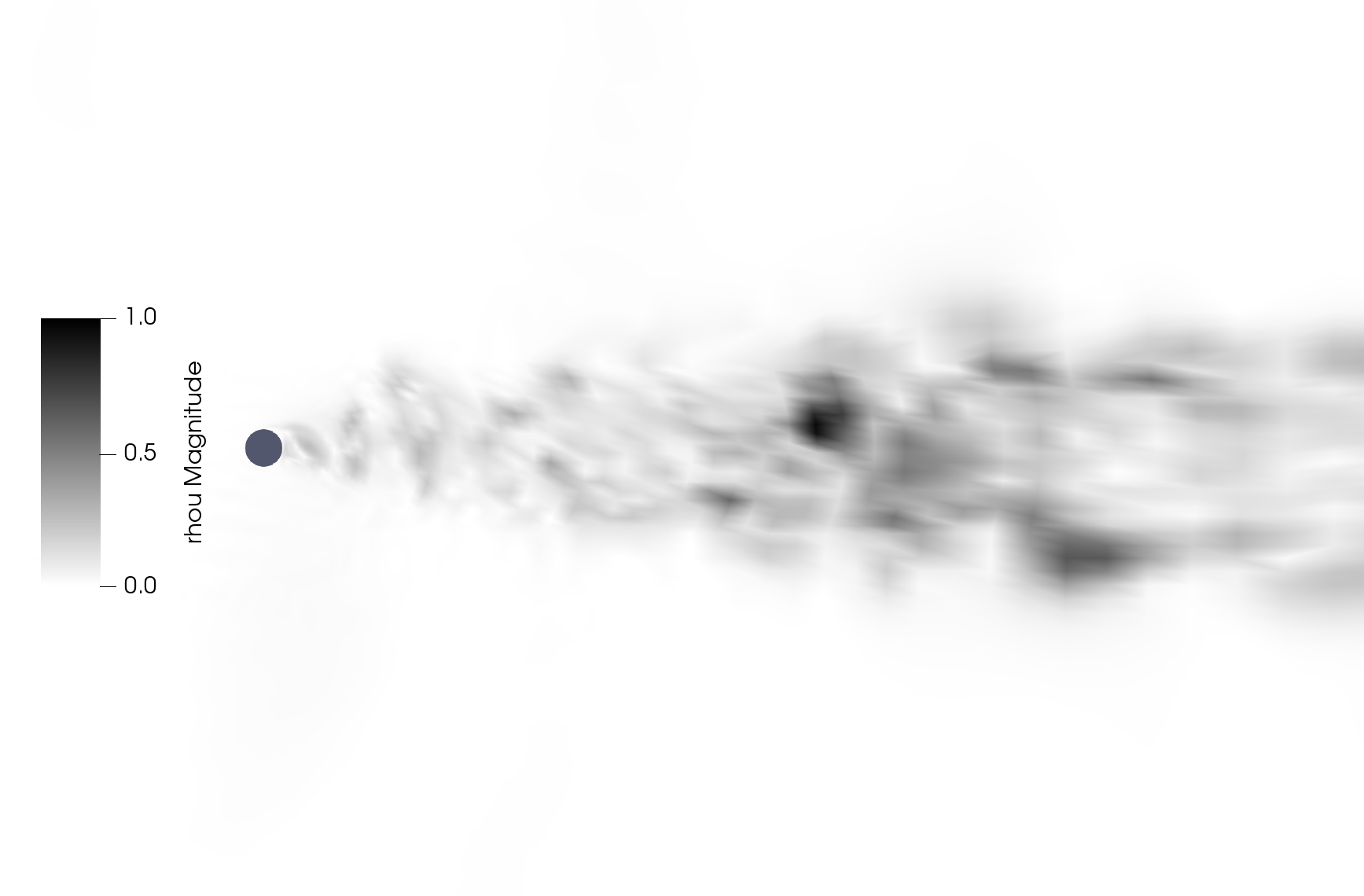}
    \includegraphics[trim=4cm 13cm 8cm 13cm, clip=true, width=0.49\textwidth]{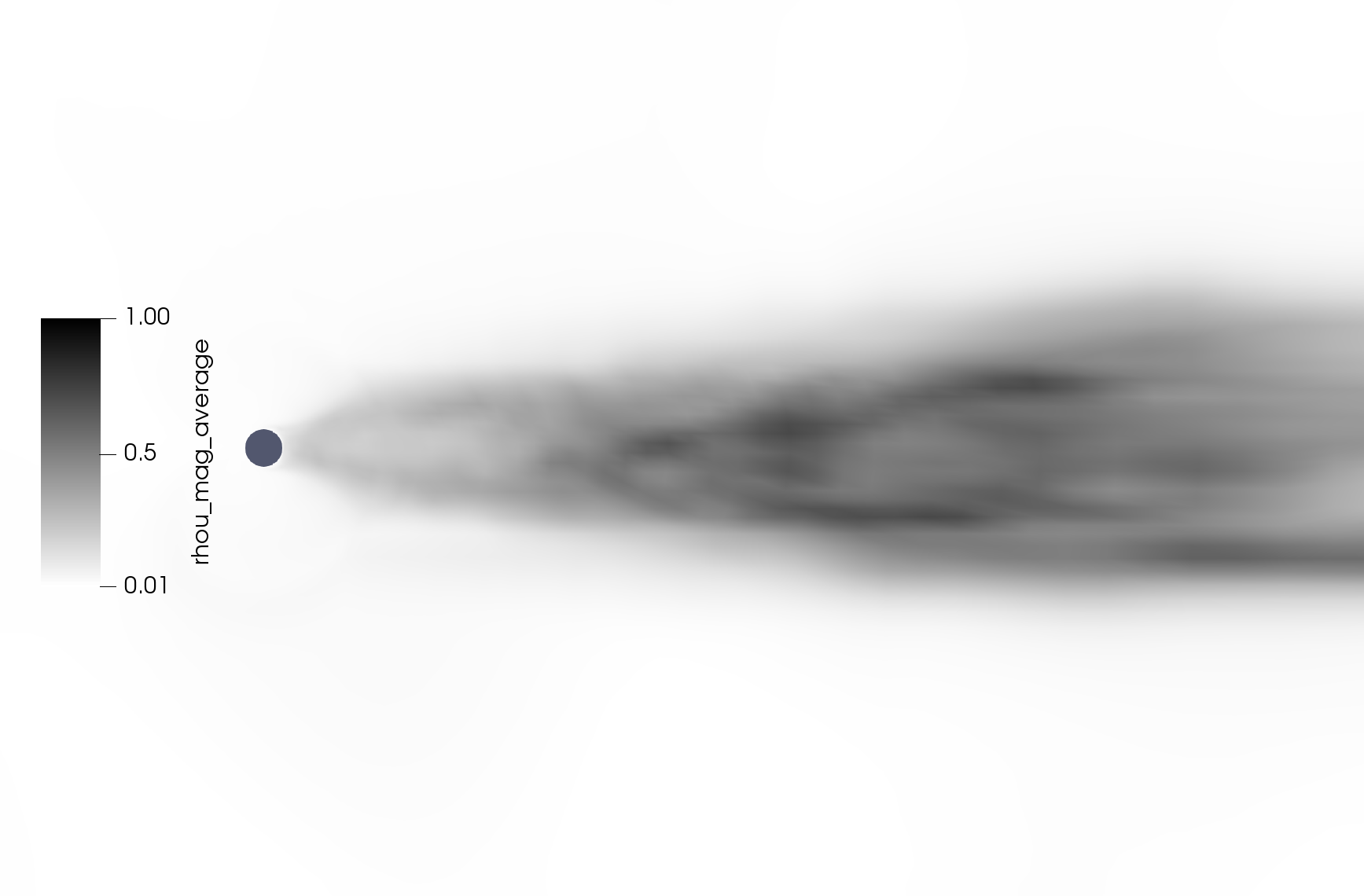}
    \caption{40th CLV}
  \end{subfigure}

  \begin{subfigure}{\textwidth}
    \includegraphics[trim=4cm 13cm 8cm 13cm, clip=true, width=0.49\textwidth]{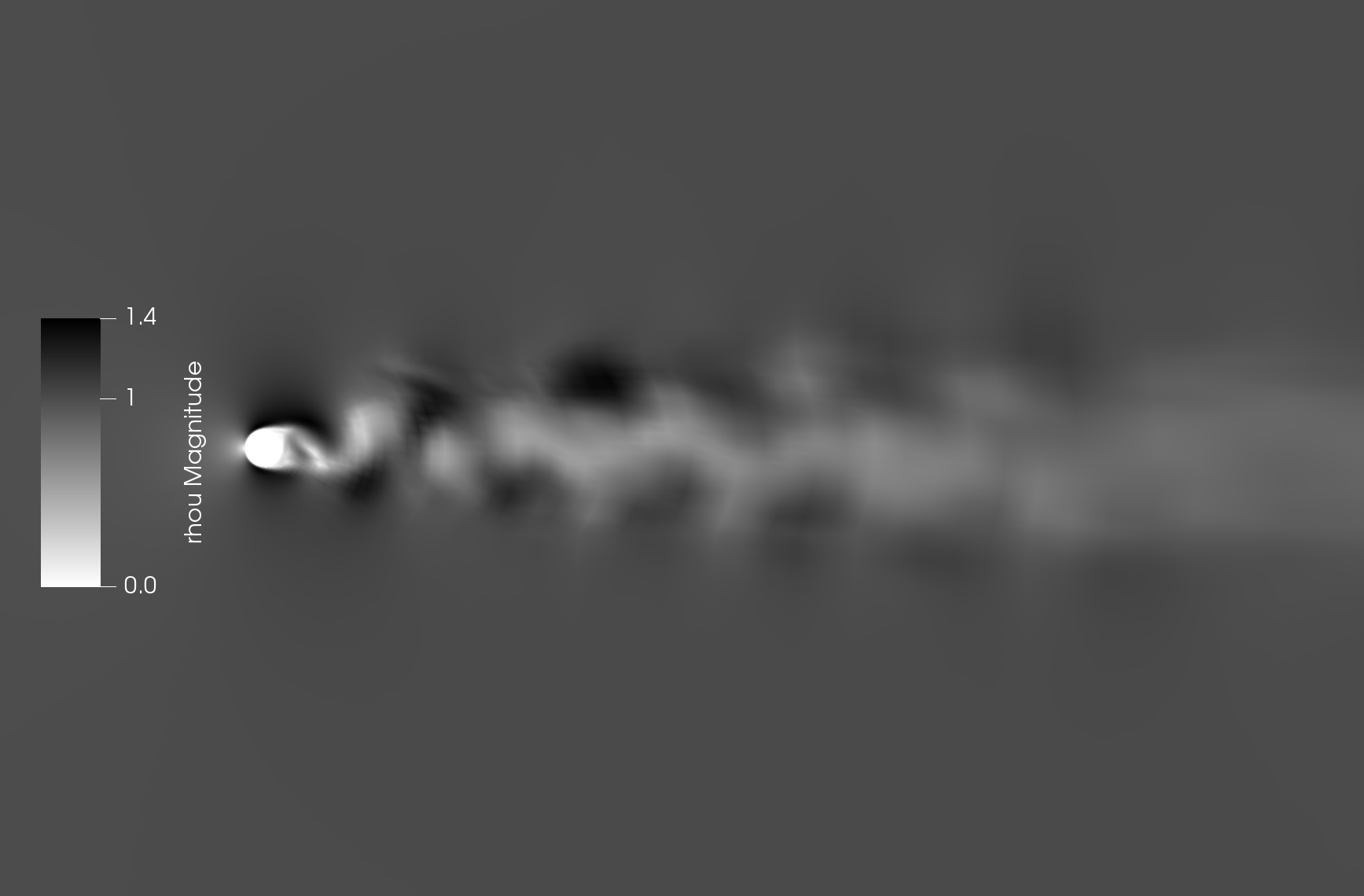}
    \includegraphics[trim=4cm 13cm 8cm 13cm, clip=true, width=0.49\textwidth]{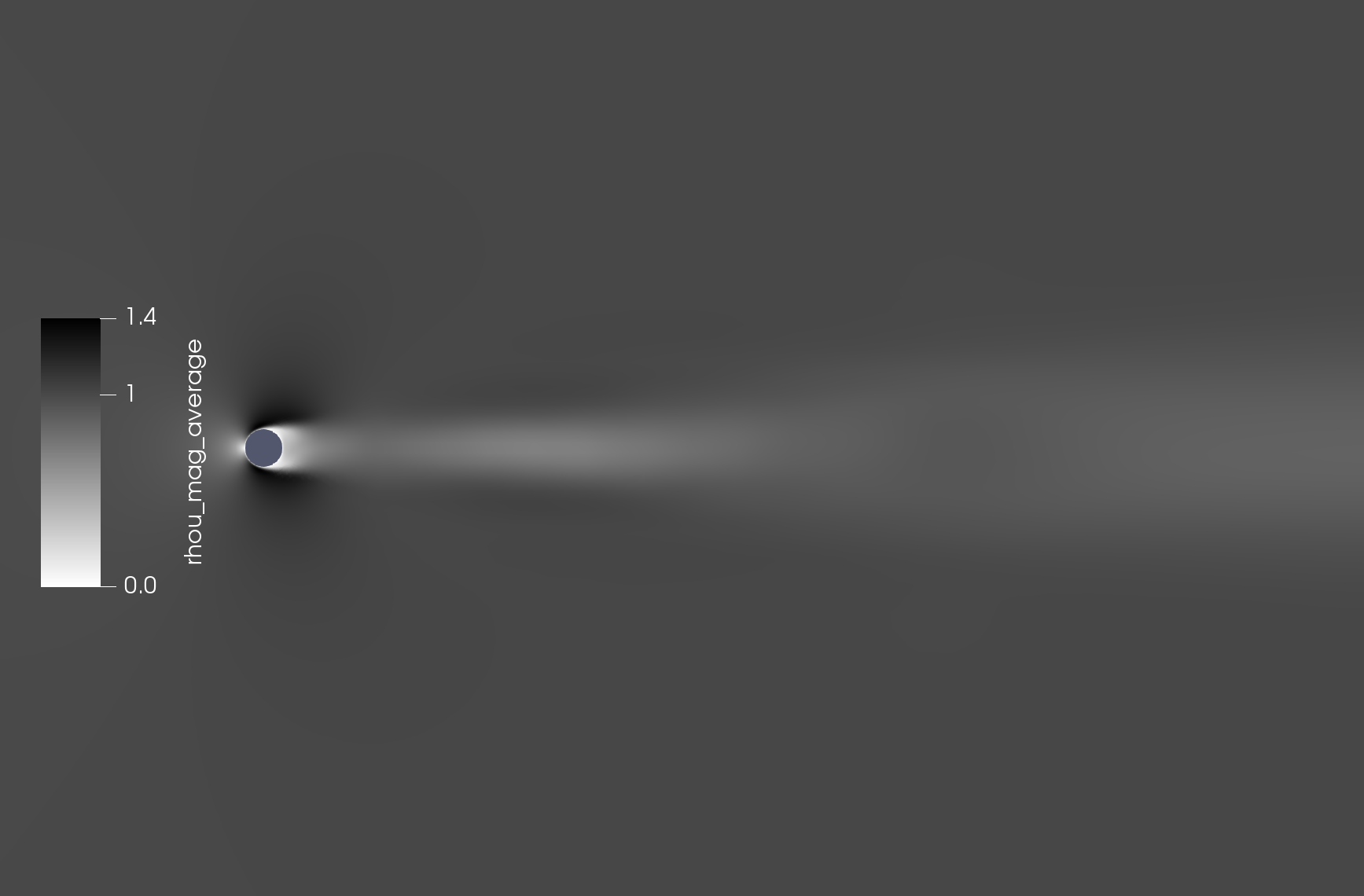}
    \caption{Primal flow field}
  \end{subfigure}

  \caption{CLV computed on the coarser mesh on a trajectory of time length $T = 211t_0$.
    Left: snapshots at $T/2$.
    Right: averaged over time span $[95t_0, 116t_0]$.
    Plotted by the magnitude of $\rho U$ component, and normalized such that the largest value is $1.00$.
    The primal flow at the bottom is normalized by $\rho_0U_0$.}
  \label{f:CLV_mid}
\end{figure}

\begin{figure}
  \centering
  \begin{subfigure}{\textwidth}
    \includegraphics[trim=4cm 13cm 8cm 13cm, clip=true, width=0.49\textwidth]{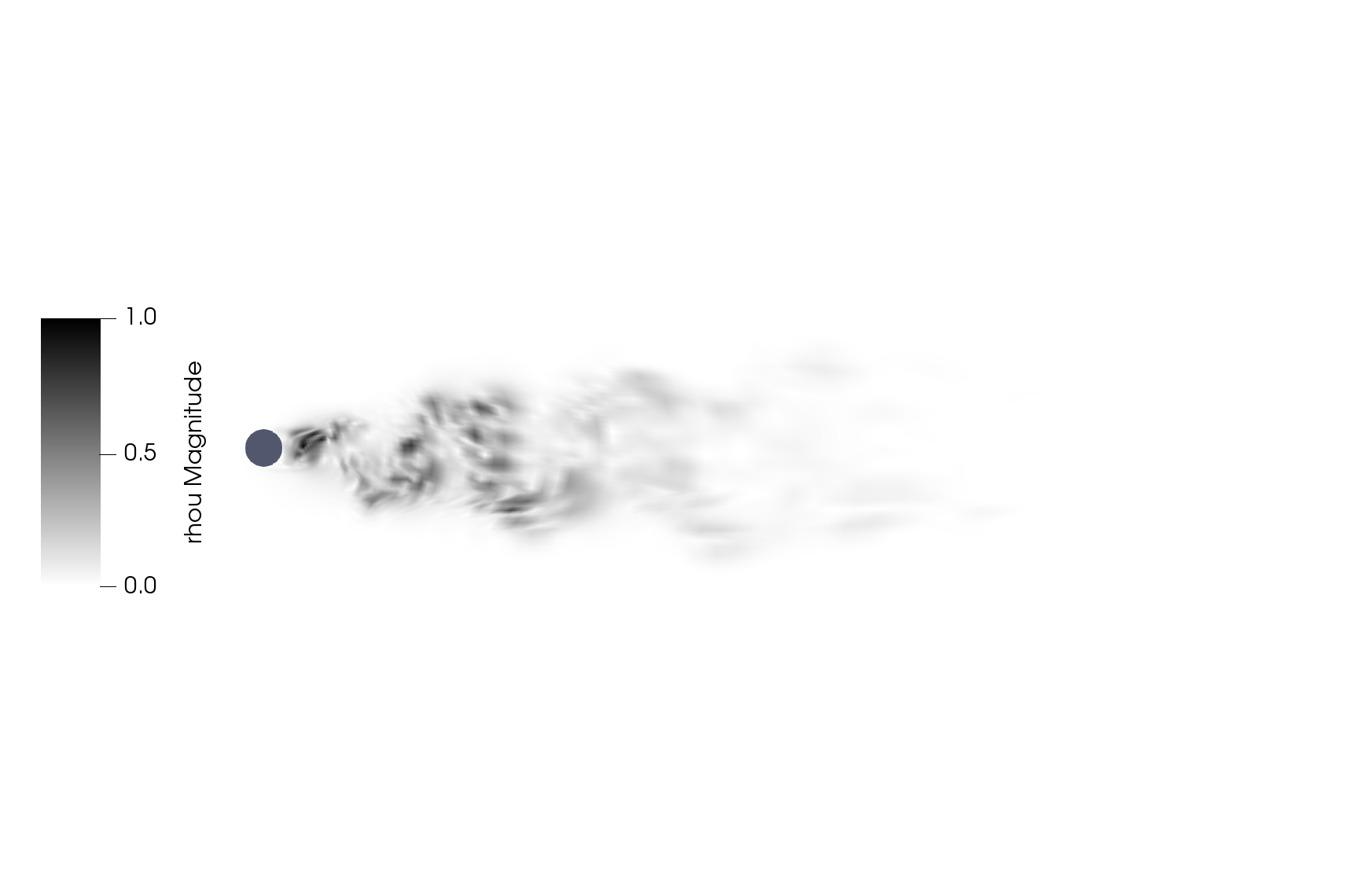}
    \includegraphics[trim=4cm 13cm 8cm 13cm, clip=true, width=0.49\textwidth]{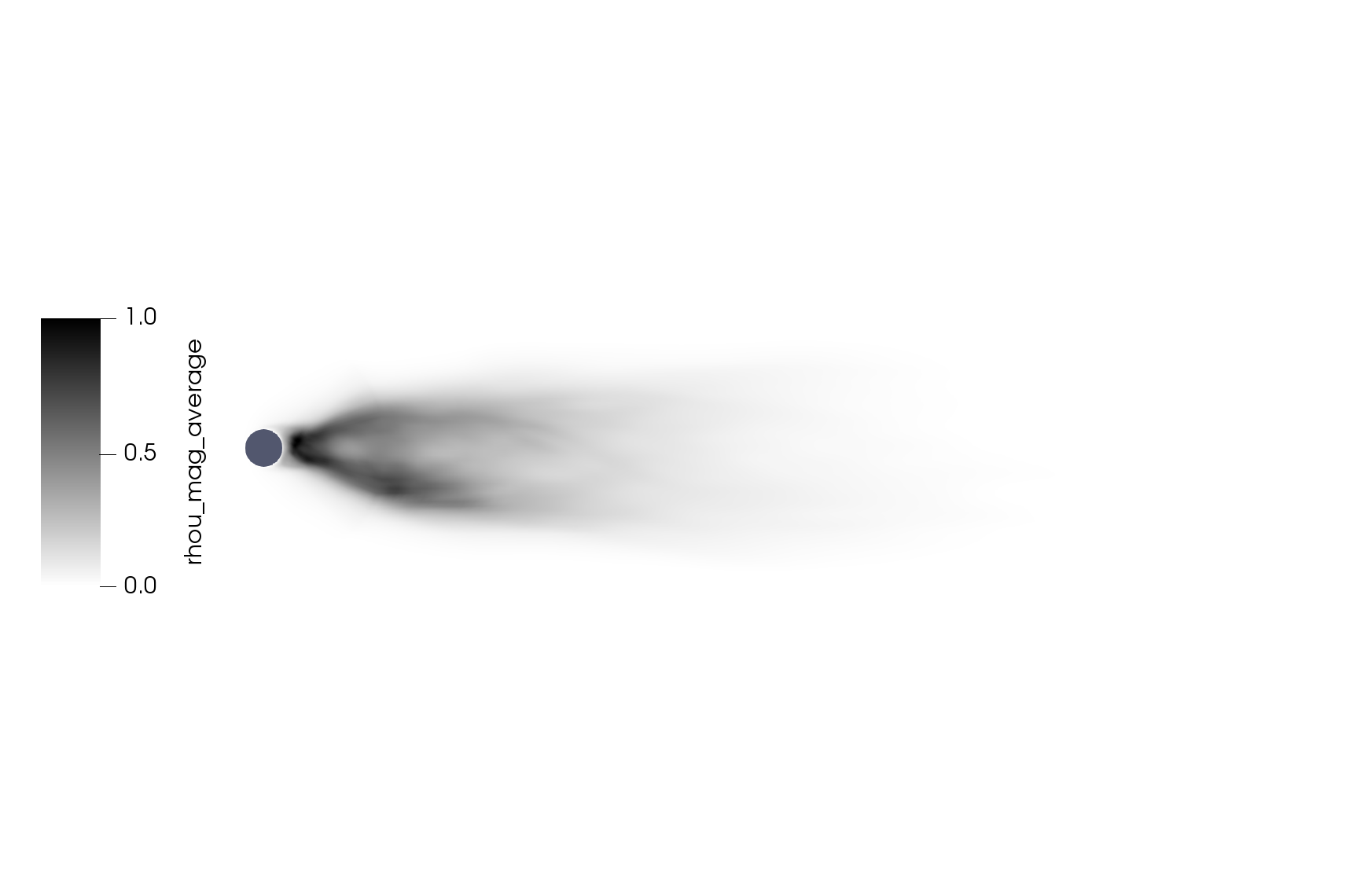}
    \caption{1st CLV}
  \end{subfigure}

  \begin{subfigure}{\textwidth}
    \includegraphics[trim=4cm 13cm 8cm 13cm, clip=true, width=0.49\textwidth]{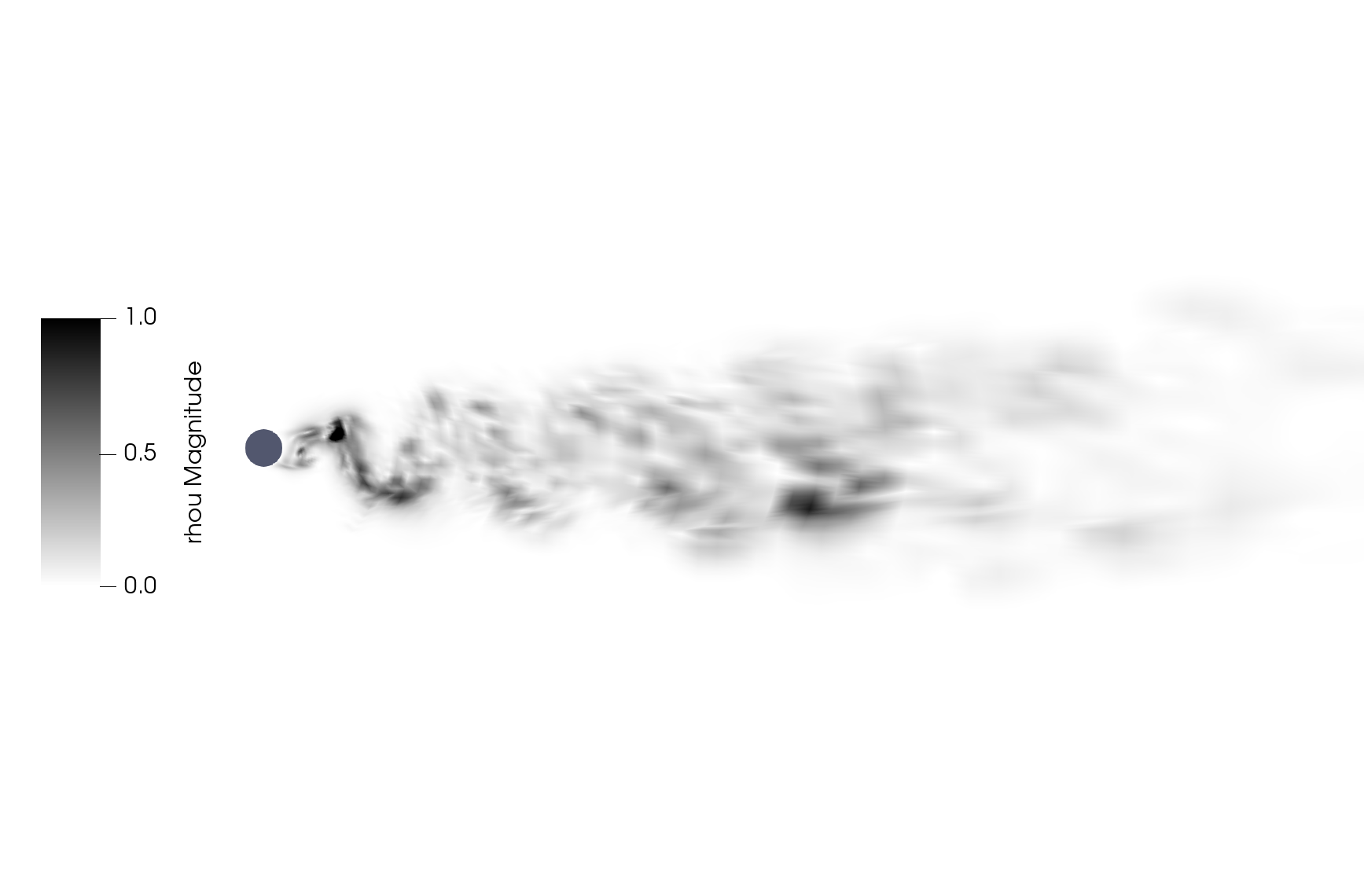}
    \includegraphics[trim=4cm 13cm 8cm 13cm, clip=true, width=0.49\textwidth]{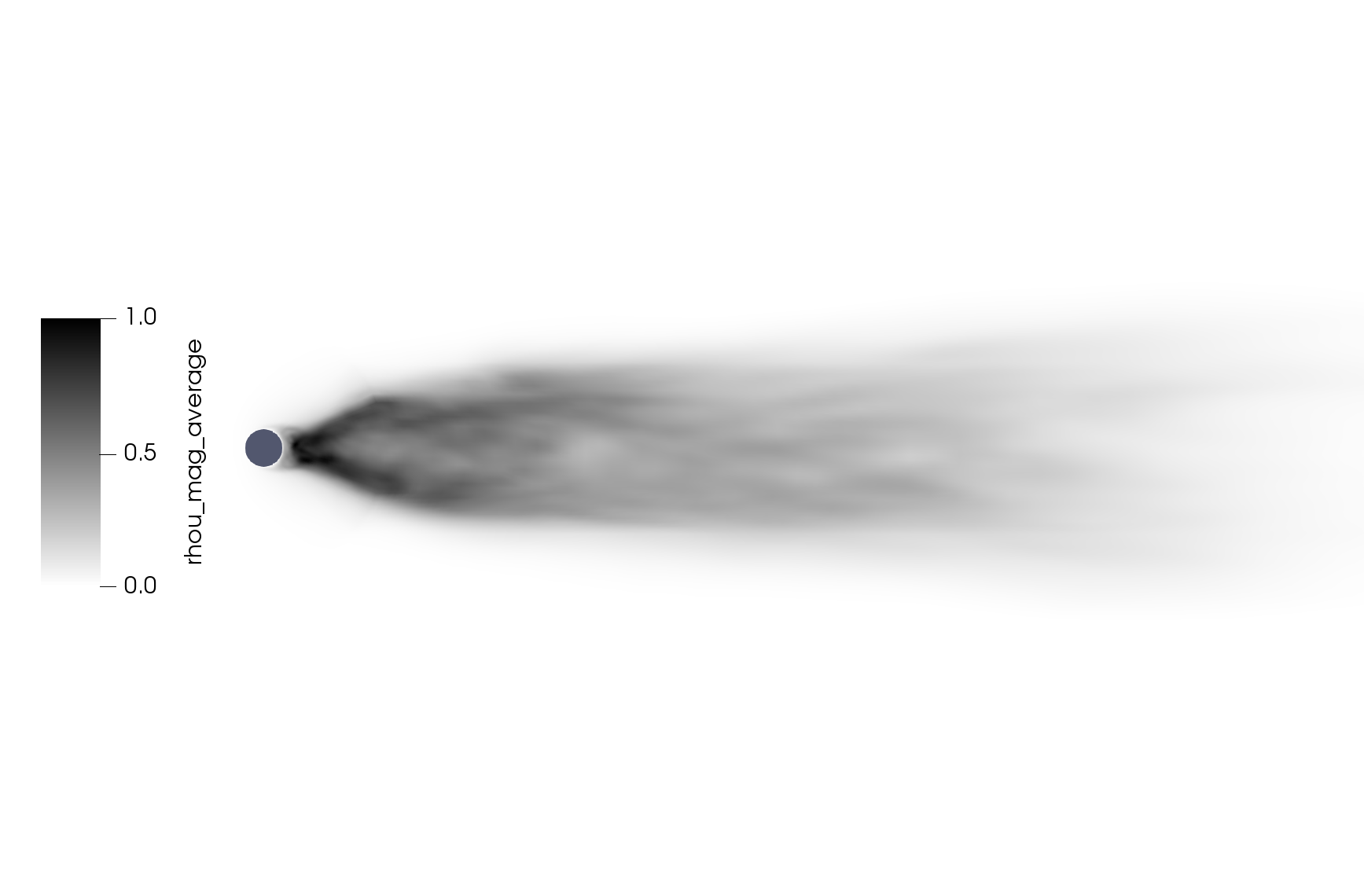}
    \caption{5th CLV}
  \end{subfigure}

  \begin{subfigure}{\textwidth}
    \includegraphics[trim=4cm 13cm 8cm 13cm, clip=true, width=0.49\textwidth]{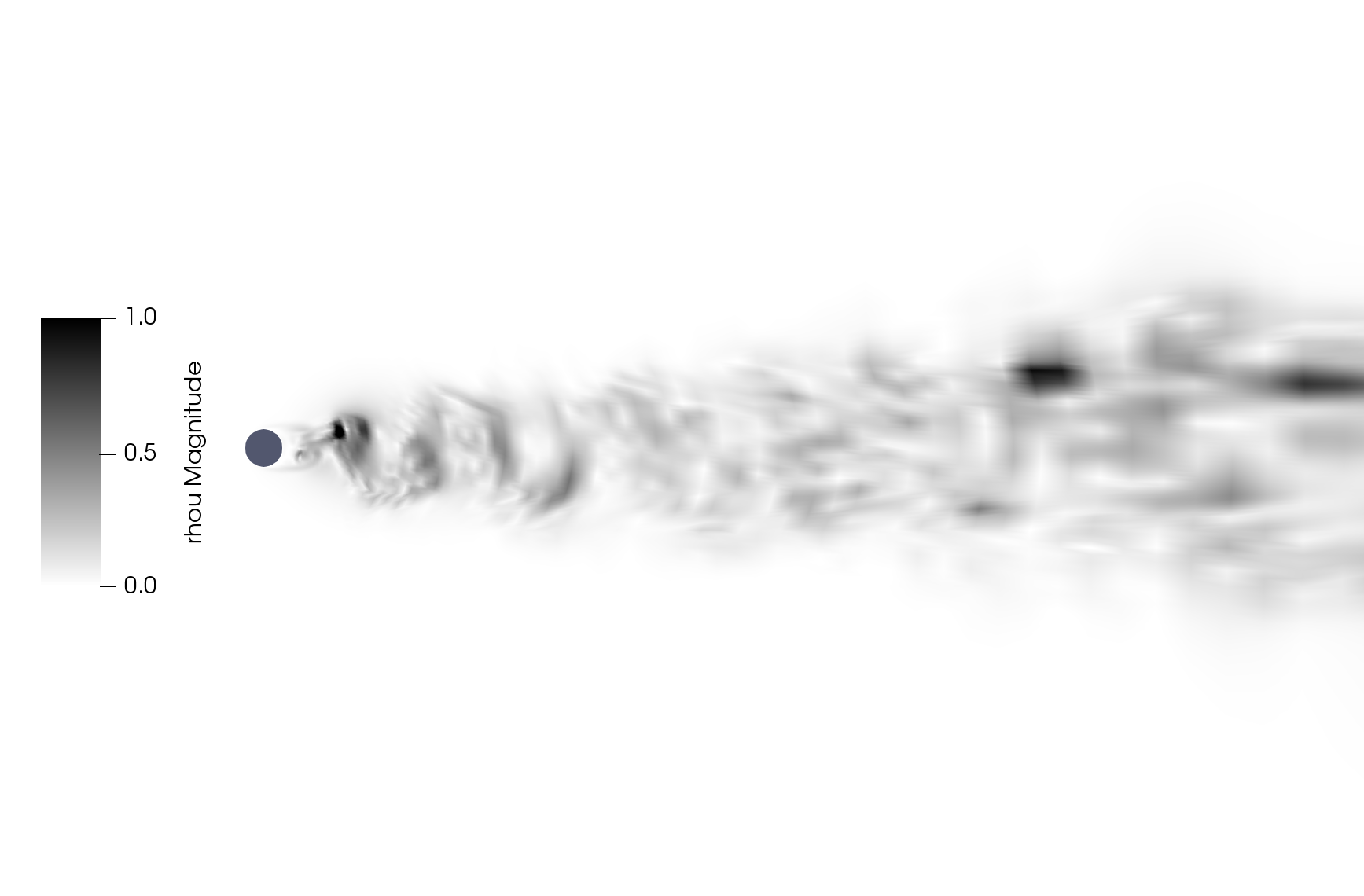}
    \includegraphics[trim=4cm 13cm 8cm 13cm, clip=true, width=0.49\textwidth]{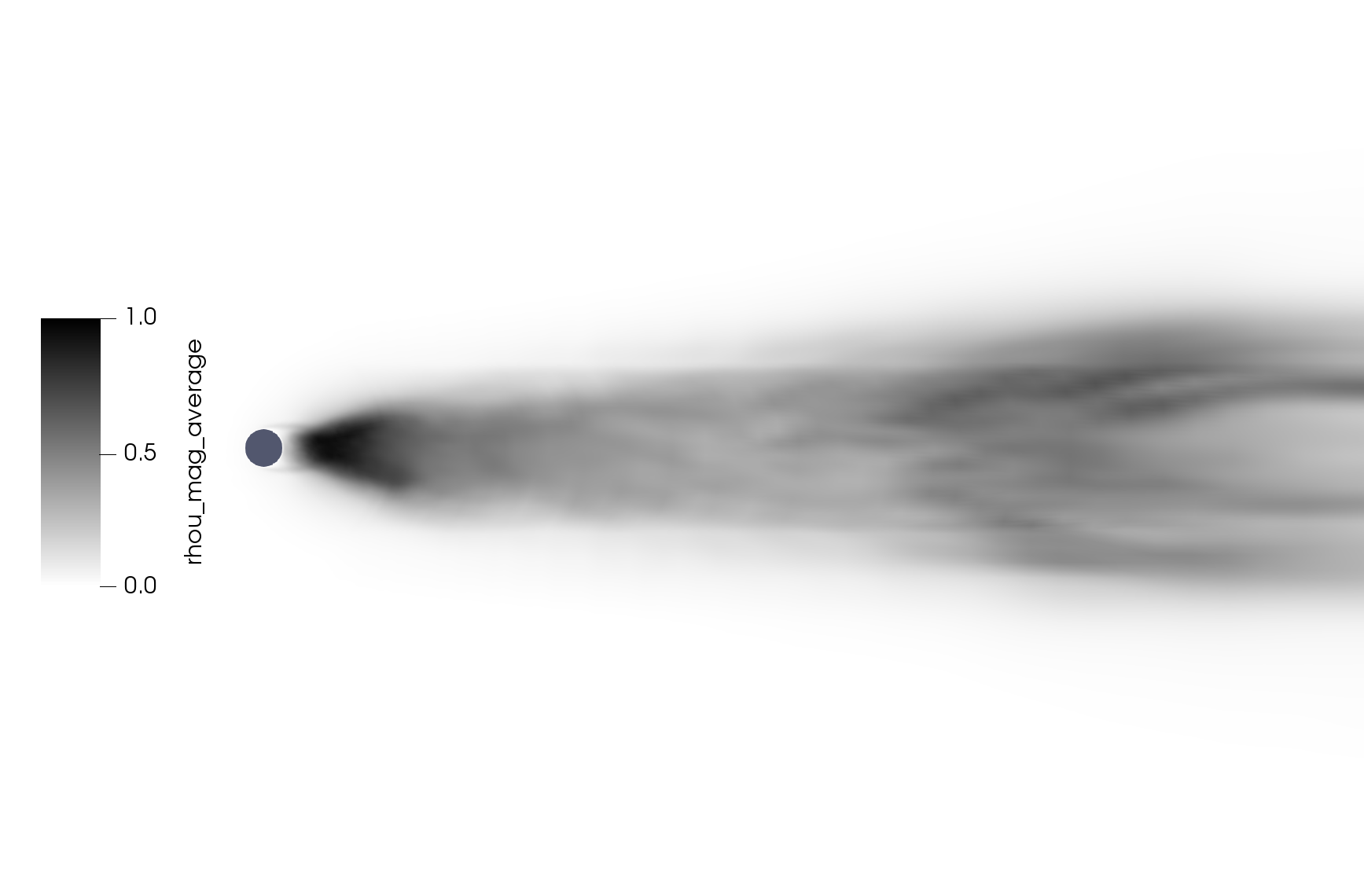}
    \caption{17th CLV}
    \label{f:neutral_CLV}
  \end{subfigure}
  
  \begin{subfigure}{\textwidth}
    \includegraphics[trim=4cm 13cm 8cm 13cm, clip=true, width=0.49\textwidth]{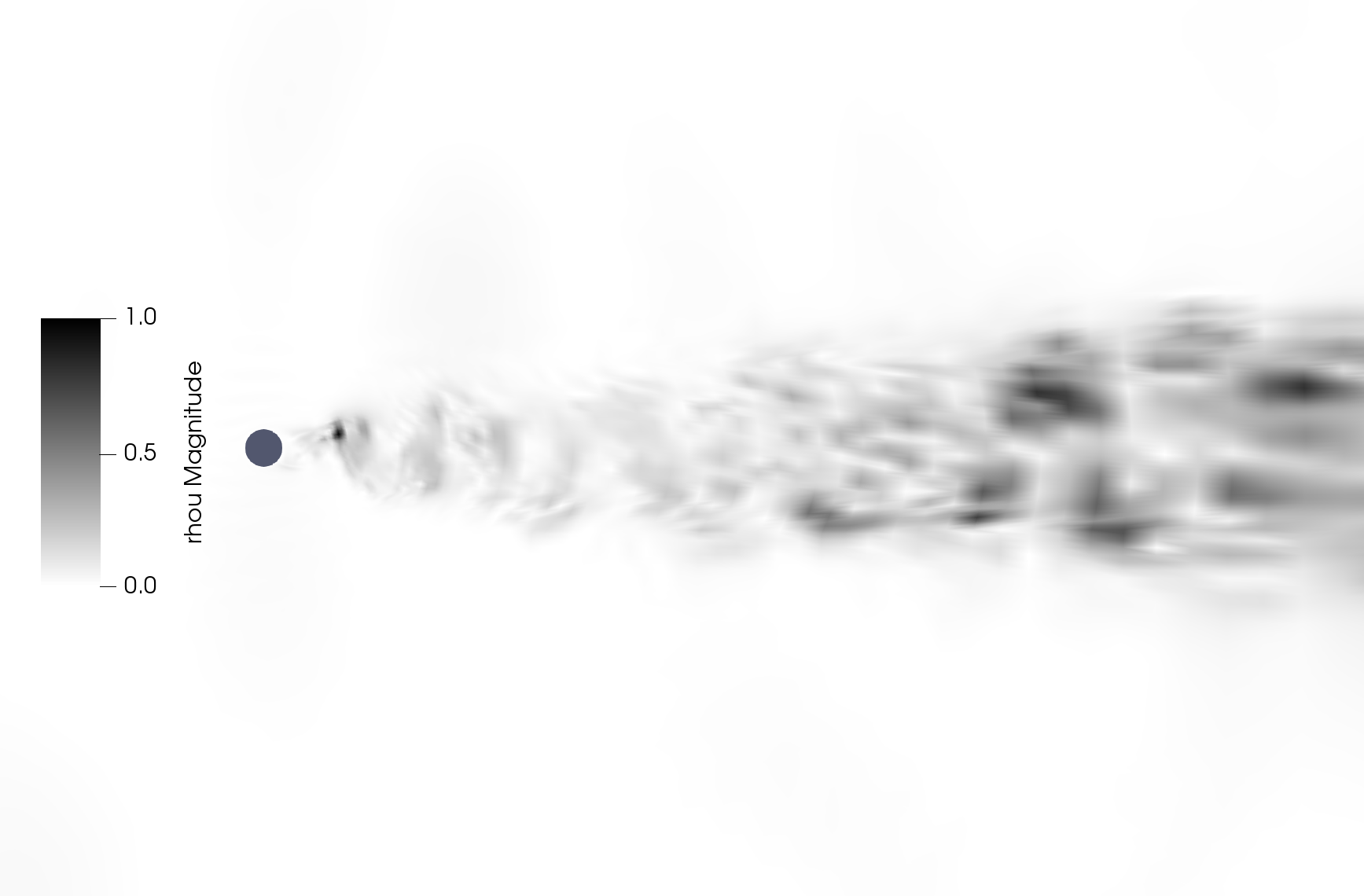}
    \includegraphics[trim=4cm 13cm 8cm 13cm, clip=true, width=0.49\textwidth]{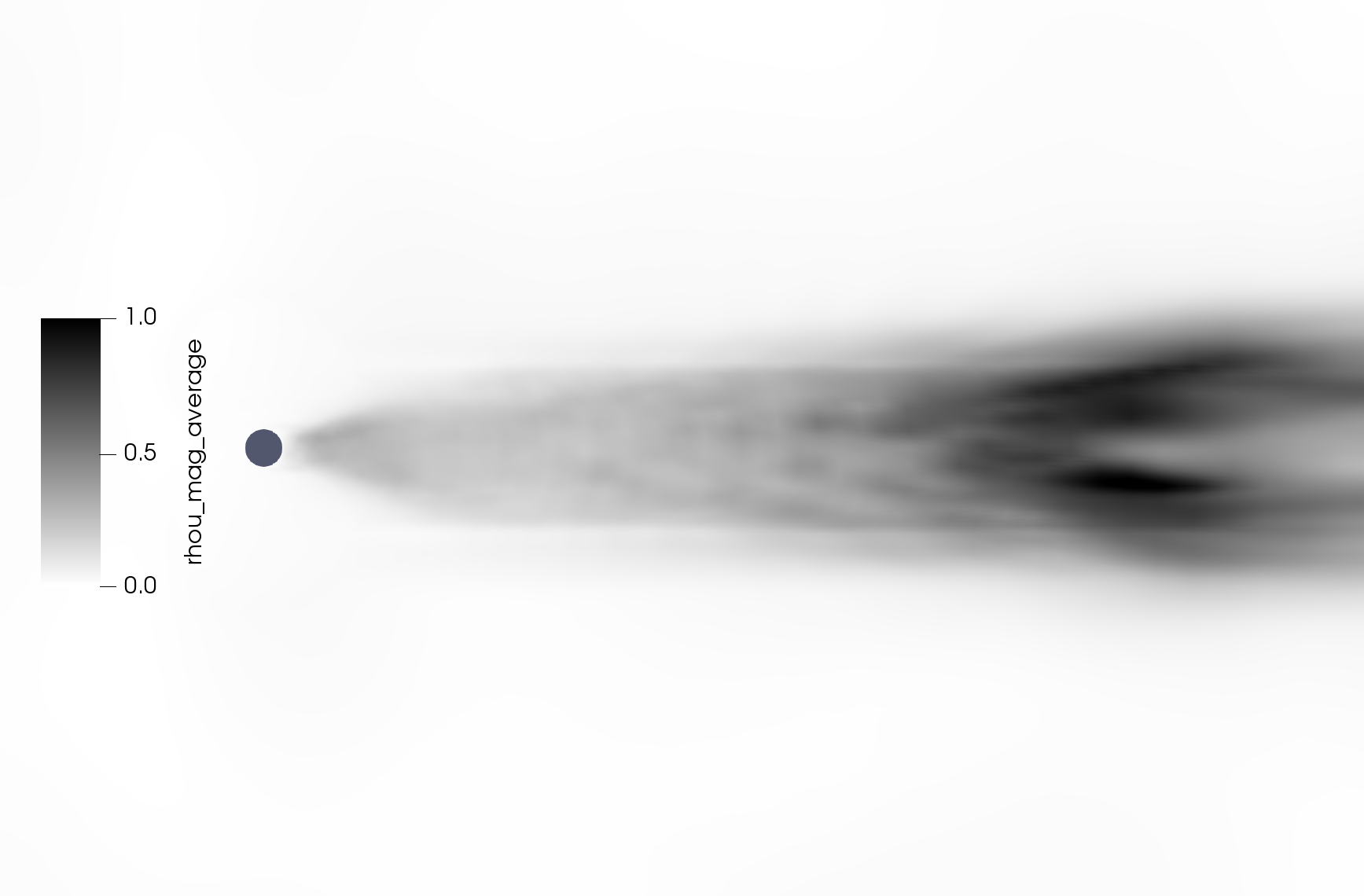}
    \caption{40th CLV}
  \end{subfigure}

  \begin{subfigure}{\textwidth}
    \includegraphics[trim=4cm 13cm 8cm 13cm, clip=true, width=0.49\textwidth]{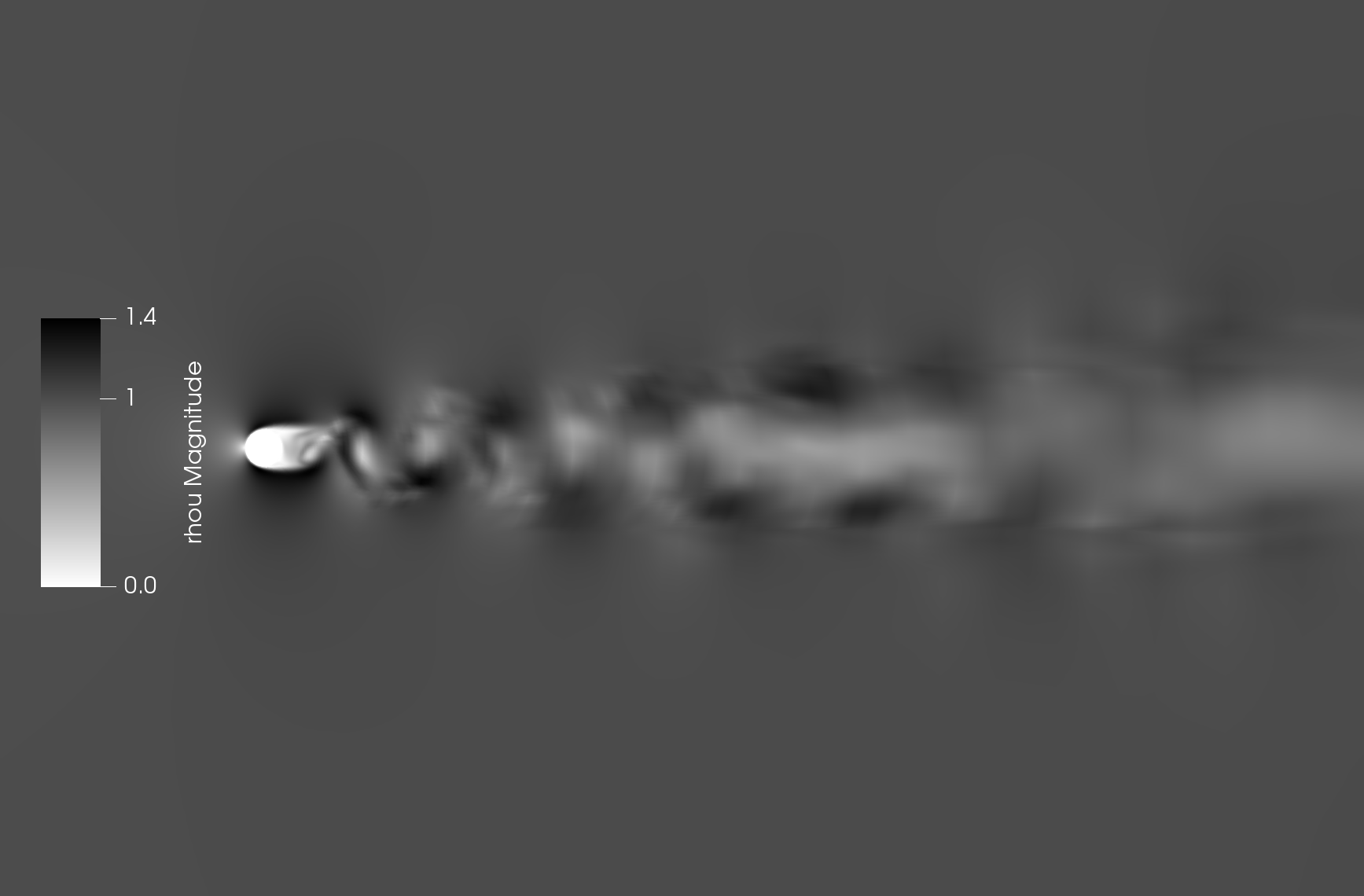}
    \includegraphics[trim=4cm 13cm 8cm 13cm, clip=true, width=0.49\textwidth]{figs/primal_averaged_finer.png}
    \caption{Primal flow field}
  \end{subfigure}

  \caption{CLV computed on the finer mesh, using same settings as the coarser mesh.}
  \label{f:CLV_finer}
\end{figure}

In fact, for two CLVs whose exponents are apart, their angles are so large that we can even observe the difference in active areas.
To illustrate, for the coarser mesh, we plot the 1st and 40th CLVs in figure~\ref{f:CLV_mid}.
For the finer mesh, we plot the 1st, 5th, 17th, and 40th CLVs in figure~\ref{f:CLV_finer}.
\footnote{Movies of the 1st, 5th, 17th and 28th CLVs on the finer mesh can be found on YouTube via
\href{https://www.youtube.com/playlist?list=PLAlGrl2Jghfu31xrP3v_rnpiRRt3RHQnR}{this link}.}
We remind readers that only the direction of CLVs, that is, only the relative flow pattern of CLVs, is meaningful.
Since our primal solution lives in the function space $X\subset\R^m$ whose tangent space is also $\R^m$, 
for any time $t$, a CLV $\zeta(t)$ lives in a function space locally the same as the primal solution $u(t)$.
In our 3D flow problem, this means our CLVs also have $\rho$, $\rho U$, $\rho E$ components, and we plot $|\rho U|$.

As we can see in both figure~\ref{f:CLV_mid} and \ref{f:CLV_finer}, the general trend is that 
the active areas of CLVs move downstream as exponents become smaller.
Intuitively, in boundary layers and near wakes, chaos is generated; perturbations in these places tend to grow, resulting in unstable CLVs.
On the other hand, in far wakes, chaos is dissipated; perturbations in this region tend to decay, resulting in stable CLVs.
Further, we can rationally conjecture that the location of neutral CLVs is in between stable and unstable CLVs,
that is, their active area is the entire wake.
This conjecture is also backed by observing the plot of a neutral CLV in figure~\ref{f:neutral_CLV}.

We conjecture that perturbations in the stagnant area in front of the cylinder will be dissipated, resulting in stable CLVs.
This is because the fluid motion in front of the cylinder is almost determined, that is, fluid here should decelerate into stagnancy.
If we make a perturbation to this area, the fluid motion in the long term will not be changed.
Hence, such a perturbation could not last long, and the corresponding LE should be negative. 
Our conjecture is also partly evidenced by the fact that the first 40 CLVs are not active in this area.

We further conjecture that perturbations in the free stream are also dissipated, resulting in stable CLVs.
We provide two arguments supporting this conjecture.
First, assume we make a perturbation in the free stream.
Then, after changing reference frame, this is equivalent to making a small perturbation in a stagnant fluid domain; 
the Euclidean norm of such a perturbation decays due to the dissipation in the Navier-Stokes equations.
Second, since our metric, induced by equation~\ref{e:define_inner_products}, is compactly supported,
a perturbation in the free stream will be transported outside the mesh where the volume weight is zero.
Hence under our norm, perturbation in the free stream decays also because of the decay in the volume weight.
In fact, because our norm of such perturbation decays to zero, the corresponding LE should be very negative.

To summarize, we make the following conjecture about active areas of CLVs in subsonic open flows.
In subsonic open flows, unstable CLVs are active in the instability-generating area such as boundary layers and near wakes;
the neutral CLV occupies similar locations as the wake;
stable CLVs with moderately negative LEs are active in the dissipative area such as far wakes;
stable CLVs active in the stagnant area ahead of the cylinder and in the free stream have very negative LEs.
Our conjecture also indicates that subsonic open flows have large angles between far-apart CLVs.

Our conjecture may provide insights into the question about how many CLVs are unstable for fluid systems.
Owing to our arguments, for open flows, or more generally fluid systems with non-turbulent regions,
there should be a large part of CLVs that are stable; 
hence unstable CLVs should be a small fraction of all CLVs.
At least for these fluid systems, more analysis should be focused on unstable CLVs, since they are the major contributor for turbulence.
Additionally, there should be more numerical methods developed which explicitly exploit the unstable structures, such as NILSS and NILSAS.

We find more structures in hyperbolicity,
but our system certainly does not satisfy the uniform hyperbolicity under which shadowing methods were mathematically developed.
Yet there is still hope, as the chaotic hypothesis suggested that, although the logical bridge of uniform hyperbolicity breaks down,
we may find tools across the river still useful.
With this hope, we go on to examine shadowing results on our fluid system.
Should we find shadowing methods still valid, we will have more confidence in its generality.

\section{Shadowing directions and sensitivities}

\subsection{Definition of shadowing directions} \label{s:define shadow}

The shadowing solution $v^\infty$ is an inhomogeneous tangent solution whose orthogonal projection perpendicular to the trajectory, 
$v^{\infty \perp}$, is uniformly bounded on an infinitely long trajectory.
Here $\cdot^\perp$ is defined as
\begin{equation}\label{e:vperp projection}
  p^\perp = p - \frac{f^T p}{f^T f} f \, ,
\end{equation}
where $p\in \R^m$ is an arbitrary vector,
$f$ is the trajectory direction defined in equation~\eqref{e:dynamical_system}, and the inner product was defined in equation~\eqref{e:define_inner_products}.
We remark here that, for a continuous dynamical system, 
the definition of shadowing direction involves a reparametrization of time \citep[Definition 1.26]{Pilyugin1999book},
or equivalently a `time dilation' \citep{Wang_ODE_LSS},
or projecting the tangent solutions onto the subspace perpendicular to $f$ \citep{Ni_NILSS_JCP}.
The $v^{\infty \perp}$ is the projection of $v^\infty$ given by the third approach, and $v^{\infty\perp}$ is bounded, but $v^\infty$ is not.
For convenience, we also call $v^\perp$ the shadowing direction.

\begin{figure}
  \centering
  \includegraphics[trim=0cm 0cm 0cm 0cm, clip=true, width=0.65\textwidth]{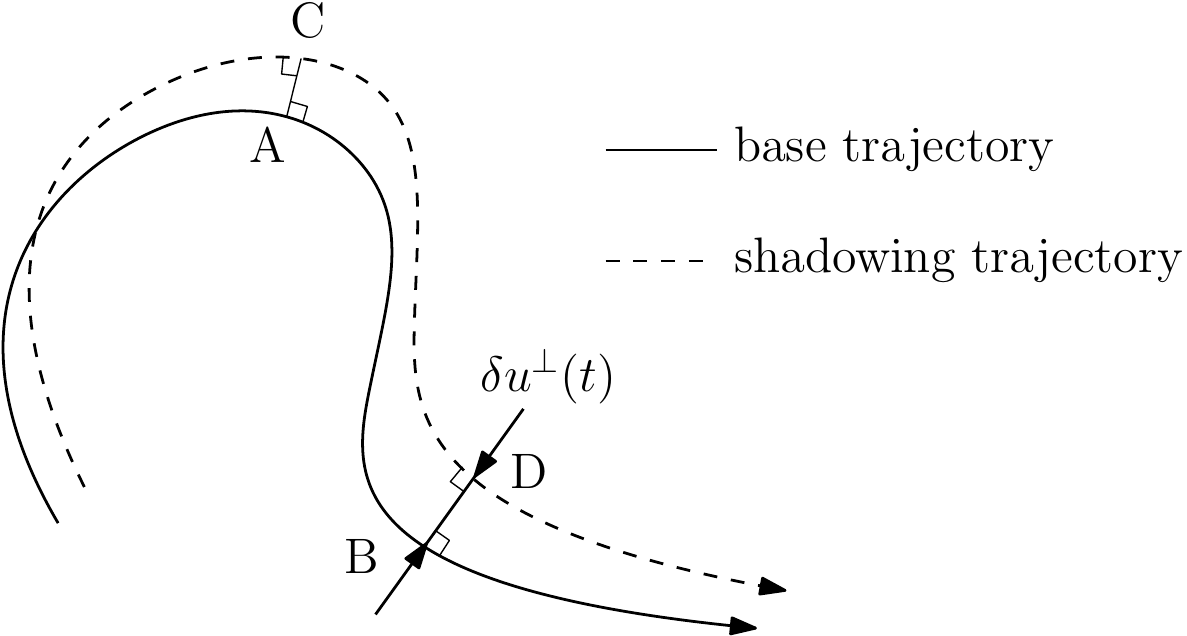}
  \caption{Shadowing directions. 
    The base trajectory has parameter $s$, the shadowing trajectory has parameter $s+\delta s$. 
    The first order approximation of the perpendicular distance is $\delta u ^ \perp \approx v^{\infty\perp}\delta s$.
    Here $v^{\infty\perp}$ is the shadowing direction, which is an uniformly bounded inhomogeneous tangent solution.}
  \label{f:shadowing_direction}
\end{figure}

The shadowing direction has the following physical meaning.
The definition of inhomogeneous tangent solution $v = du/ds$ tells us that, with perturbed parameter $s+\delta s$, 
there is a new trajectory $u+\delta u$ such that $\delta u \approx v^\infty \delta s$.
Now the perpendicular distance between the new and the old trajectory is $\delta u ^\perp \approx v^{\infty\perp} \delta s$;
together with the boundedness of $v^{\infty \perp}$,
we see that the new trajectory remains perpendicularly close to the old trajectory for a long time.
This new trajectory will be referred to as the shadowing trajectory, and the old one as the base trajectory.
This intuition of the shadowing direction is shown in figure~\ref{f:shadowing_direction}.

The initial condition of the shadowing direction is not known \textit{a priori}, 
or equivalently, we do not specify how system parameters would affect initial conditions.
In fact, as we shall see, although the shadowing direction solves that same ODE as in equation~\eqref{e:inhomo tangent},
its determining condition is no longer the initial condition;
rather, to determine a shadowing direction, we minimize the $L^2$ norm of the orthogonal projection of an inhomogeneous tangent solution.

\subsection{Sensitivity analysis of long-time-averages via shadowing directions}

For chaotic dynamical systems such as the 3D flow problem in this paper, the output of the system, such as the drag or lift, is typically aperiodic.
As often happens in engineering, the objective is the long-time-average $\avg{J}_\infty$ of an instantaneous quantity $J(u,s):\R^m\times\R\rightarrow\R$.
More specifically, we define
\begin{equation} \label{e:average J}
  \avg{J}_\infty:= \lim\limits_{T\rightarrow\infty}\avg{J}_T, \text{ where }\avg{J}_T:= \frac{1}{T}\integrate J(u,s) dt.
\end{equation}
Typically, $\avg{J}_\infty$ is approximated by a $\avg{J}_T$ with a large $T$.
The sensitivity of the objective with respect to system parameters, $d\avg{J}_\infty/ds$, is of engineering interest.
In this subsection we show how to compute the sensitivity using the shadowing direction.

Assume we have found, on a finitely long trajectory, an approximation $v\approx v^\infty$, where $v$ is also an inhomogeneous tangent solution.
We first define the time dilation term \citep{Wang_ODE_LSS} $\eta$:
\begin{equation} \label{e:eta and v}
  \dd{v^\perp}{t} = \partial_u f v^\perp + \partial_s f + \eta f\,.
\end{equation}
Intuitively, $\eta$ describes the relative time spent on the shadowing trajectory in comparison with the base trajectory.
In figure~\ref{f:shadowing_direction}, if
the shadowing trajectory takes less time to travel from point C to D than the base trajectory from A to B, 
then equation~(\ref{e:eta and v}) gives $\eta < 0$.
On the other hand, if the shadowing trajectory moves slower, then $\eta>0$.

Once we get $v$ and $\eta$, the sensitivity can be computed by
\begin{equation} \label{e:djds with eta}
  \dd{\avg{J}_\infty}{s} \approx 
  \frac 1T \int_{0}^{T} \left[\partial_u J \, v^{\perp}+ \partial_sJ + \eta (J - \avg{J}_T) \right] \, dt \,,
\end{equation}
where $\avg{J}_T$ is defined in equation~(\ref{e:average J}). 
Intuitively, the first term inside the integration in equation~(\ref{e:djds with eta}) 
describes the contribution by the perpendicular distance between the shadowing trajectory and the base trajectory.
The second term is due to the fact that the function $J$ may explicitly depend on $s$.
The last term involves $\eta$, and it accounts for the fact that the average is taken with respect to time:
if in some small region the shadowing trajectory moves faster than the base trajectory, 
then the time spent is shorter, hence the contribution from this region in the time average should also be smaller.

Another formula for the sensitivity is easier for computer programming: 
\begin{equation} \label{e:djds with xi}
  \dd{\avg{J}_\infty}{s} \approx 
  \frac 1T \left[
  \int_{0}^{T} \left(\partial_u J \, v+ \partial_sJ \right)dt 
  +\left. \xi\right\vert^T_0 \avg{J}_T
  -\left. \left(\xi J \right)\right\vert^T_0\right] \,,
\end{equation} 
where the time difference term, $\xi$, is a time-dependent scalar function such that
\begin{equation} \label{e:xi}
  \xi f = v-v^\perp \,.
\end{equation}
Thus $\xi$ is easier to compute than $\eta$, since its definition does not involve a time derivative.
Note that in \eqref{e:djds with xi} we use $v$ instead of its projection $v^\perp$.
The derivation of \eqref{e:djds with eta}, \eqref{e:djds with xi}, and their relation can be found in the appendix of \citep{Ni_NILSS_JCP}.

\subsection{The NILSS algorithm for computing shadowing directions}

We describe the non-intrusive least-squares shadowing (NILSS) algorithm in this subsection.
As in section~\ref{s:algorithm_CLV}, for $i=0, \dots, K-1$, we define the $i$-th time segment as $[t_i, t_{i+1}]$, with $t_i=i\Delta T$.
In the algorithm presented below, for quantities defined on the entire segments, such as $u_i, W_i, v^*_i$, $C_i$, and $d_i$, 
we use the same subscripts as the segments they are defined on.
For quantities defined only at interfaces between segments, such as $Q_{i}, R_{i}$, and $b_{i}$, 
we use the same subscripts as the time points they are defined at.
Again, we use finite difference results to approximate tangent solutions:
such a variant is called the finite difference non-intrusive least-squares shadowing (FD-NILSS) algorithm, 
whose details are given in \citep{Ni_fdNILSS}.

To start with, we should prescribe:
(1) number of homogeneous tangent solutions, $M$, which must be larger than the number of unstable CLVs, $\mus$; 
(2) length of each time segment, $\Delta T$; 
and (3) number of time segments, $K$.
Consequently, the time length of the entire trajectory, $T=K\Delta T$ is also determined.
Then, the NILSS algorithm is given by the following procedure.

\begin{enumerate}
  \item Generate initial conditions for primal solutions, homogeneous tangent solutions, and inhomogeneous tangent solutions.

  \begin{enumerate}
    \item Compute the primal solution of \eqref{e:dynamical_system} for sufficiently long time so that the trajectory lands on the attractor, 
      then set $t=0$, and set the initial condition of the primal system, $u_0(0)$.

    \item Randomly generate an $m\times M$ orthogonal matrix $Q_{0}= [ q_{01}, \dots, q_{0M} ]$ whose column vectors are orthogonal to $f(t=0)$.
      This $Q_0$ will be used as initial conditions for homogeneous tangent solutions.

    \item Set the initial condition of the particular inhomogeneous tangent solution $v_0^{*}(0)=0$.
  \end{enumerate}

  \item For $i=0$ to $K-1$, on segment $i$, where $t\in[t_i,t_{i+1}]$, do:

  \begin{enumerate}
    \item Compute the primal solution $u_i(t)$ from $t_i$ to $t_{i+1}$.

    \item Compute homogeneous tangent solutions $W_i(t) = [ w_{i1}(t), \dots, w_{iM}(t) ]$.

    \begin{enumerate}
      \item For each homogeneous tangent solution $w_{ij}$, $j=1, \dots, M$, starting from initial condition $w_{ij}(t_i) = q_{ij}$,
        integrate equation~\eqref{e:homo tangent} from $t_i$ to $t_{i+1}$.

      \item Compute orthogonal projection $W_i^\perp(t)= [ w_{i1}^\perp(t), \dots, w_{iM}^\perp(t) ]$ via equation~(\ref{e:vperp projection}).

      \item Compute and store the covariant matrix on segment $i$:
        \begin{equation} \label{eq:covariant matrix}
          C_i = \int_{t_{i}}^{t_{i+1}} (W_i^\perp)^T W_i^\perp dt. 
        \end{equation}

      \item Perform QR factorization: 
        $W_i^\perp(t_{i+1}) = Q_{i+1} R_{i+1}$, where $Q_{i+1} = [q_{i+1,1}, \dots, q_{i+1,M}]$.
    \end{enumerate}

  \item Compute the particular inhomogeneous tangent solution $v_i^*(t)$.

    \begin{enumerate}
      \item Starting from initial condition $v^*_i(t_{i})$, 
        integrate the inhomogeneous equation~\eqref{e:inhomo tangent} from $t_i$ to $t_{i+1}$. 

      \item Compute the orthogonal projection $v_i^{*\perp}(t)$ via equation~(\ref{e:vperp projection}).

      \item Compute and store
        \begin{equation} 
          d_i = \int_{t_{i}}^{t_{i+1}} {W_i^\perp}^T v^{*\perp}_i dt .
        \end{equation}

      \item Orthogonalize $v^{*\perp}_i(t_{i+1})$ with respect to $W^{\perp}_{i+1}(t_{i+1}) = Q_{i+1}$ to obtain the initial condition of the next time segment:
        \begin{equation} \label{eq:v* initial}
          v^*_{i+1}(t_{i+1}) = v^{*\perp}_i(t_{i+1}) - Q_{i+1} b_{i+1} ,
        \end{equation}
        compute and store
        \begin{equation} 
          b_{i+1} = Q_{i+1}^T v^{*\perp}_i(t_{i+1}) \,. 
        \end{equation}
    \end{enumerate}
    
  \end{enumerate}

  \item Solve the NILSS problem:
    \begin{equation} \begin{split} \label{e:NILSS on multiple segments} 
      &\min_{\{a_i\}} \sum_{i=0}^{K-1} \frac{1}{2} a_i^T C_i a_i + d_i^T a_i \\
      \mbox{s.t. }& 
      a_{i} = R_{i} a_{i-1} + b_{i} \quad i=1,\dots,K-1.
    \end{split}\end{equation}
  This is a least-squares problem with arguments $\{a_i\}_{i=0}^{i=K-1}$, where $a_i\in \R^M$ for each $i$.

  \item Compute $ v_i $ within each time segment $t\in[t_i,t_{i+1}]$:
    \begin{equation} \label{e:v_i}
      v_i(t) = v^*_i(t) + W_i(t) a_i .
    \end{equation}
    The orthogonal projections of $v_i$ on each segment, $\{v_i^\perp\}_{i=0}^{K-1}$,
    converge to the orthogonal projection of the shadowing direction, $v^{\infty\perp}$, as both $t$ and $T-t$ become large.
    In other words, we have smaller error in the middle section of a long time span.

  \item Compute $\xi_i$ at the end of each segment:
    \begin{equation} \label{e:xi end value}
      \xi_i(t_{i+1}) = \frac{( v_i(t_{i+1}) )^T f(u(t_{i+1}))}{f(u(t_{i+1}))^T f(u(t_{i+1}))} \,.
    \end{equation}  

  \item The derivative can be computed by
    \begin{equation} \label{e:djds seg}
      \frac{d\avg{J}_\infty}{ds} \approx
      \frac 1T\sum_{i=0}^{K-1} \left[
      \int_{t_i}^{t_{i+1}} 
      \left(\partial_u J \, v_i+ \partial_sJ \right)dt 
      + \xi_i(t_{i+1}) (\avg{J}_T - J(t_{i+1})) \right] .
    \end{equation}
\end{enumerate}

Readers should note that, although at the beginning of each segment, $v_i$ is perpendicular to $f(t_i)$,
typically $v_i$ is not perpendicular to $f(t_i)$ for the rest of the segment.
In NILSS, we typically use perpendicularly projected vectors, such as $v_i^{*\perp}$, $W_i^\perp$, $v_i^\perp$, 
for constructing the minimization problem in equation~\eqref{e:NILSS on multiple segments} to compute the shadowing directions.
Unprojected vectors, such as $v_i$, are typically used for computing sensitivities by equation~\eqref{e:djds seg}.
As shown in the appendix of \citep{Ni_NILSS_JCP}, we can construct a continuous $v$, 
and its orthogonal projection $v^\perp$ confined on each segment yields $v_i^\perp$.
However, $v_i$ are not confinements of $v$, and $v_i$ are not necessarily continuous across segments.
In fact, in equation~\eqref{e:NILSS on multiple segments}, the minimization is minimizing $\|v^\perp\|_{L^2}$,
while the constraint prescribes that $v^\perp$ is continuous across all time segments.

Intuitively, at $t=0$, the difference between $v^{\infty\perp}$ and $v^{\perp}$ contains unstable CLVs;
since $v^{\infty\perp}$ is bounded, as time evolves, 
the exponential growth of unstable CLVs increases both $\|v^{\infty\perp}-v^{\perp}\|_{L^2}$ and $\|v^{*\perp}\|_{L^2}$.
Also, as time evolves, the span of $M$ homogeneous tangent solutions approximates the span of the first $M$ CLVs,
hence $v^\perp = v^{*\perp} + W^\perp a$ allows us to subtract unstable CLVs from $v^{\perp}$.
Now minimizing $\|v^{\perp}\|_{L^2}$ removes unstable CLVs from both $v^{\perp}$ and $v^{\infty\perp}-v^{\perp}$,
thus making $v_i^\perp \approx v^{\infty\perp}$.
Since the number of unstable CLVs is typically much smaller than the dimension of the dynamical system, NILSS is computationally efficient.
A more detailed explanation of NILSS is contained in \citep{Ni_NILSS_JCP},
and explanation of FD-NILSS is contained in \citep{Ni_fdNILSS}.
\footnote{The python package `fds' implementing FD-NILSS is available on GitHub via \href{https://github.com/qiqi/fds}{this link}.
  The particular files related to the application in this section are in fds/apps/charles\_cylinder3D.
  Note this is a different folder from that for the last section.}

\subsection{Results of shadowing directions}

This subsection discusses the shadowing directions computed by the FD-NILSS algorithm.
We remind readers that shadowing solutions depend only on the choice of system parameters but not on objectives.
In particular, this subsection shows two shadowing solutions with respect to two system parameters:
(1) free-stream velocity $U$, normalized by $U_0$;
and (2) rotation speed of the cylinder $\omega$, measured in cycles per time unit, normalized by $\omega_0$.

In FD-NILSS we set time-step size $1\times10 ^ {-8}$, and $200$ steps in each time segment;
hence the segment length is $\Delta T = 2\times 10 ^{-6} = 0.259 t_0$.
These are the same as what we used for computing LEs and CLVs.
Different from those used for computing LEs and CLVs, we set the number of homogeneous tangents $M=30$ and the number of segments $K=600$;
hence the time length of the entire trajectory is $T = 1.2\times 10^{-3} = 158.4 t_0$.

\begin{figure}
  \centering
  \begin{subfigure}{0.49\textwidth}
    \includegraphics[trim=0cm 0cm 0cm 0cm, clip=true, width=\textwidth]{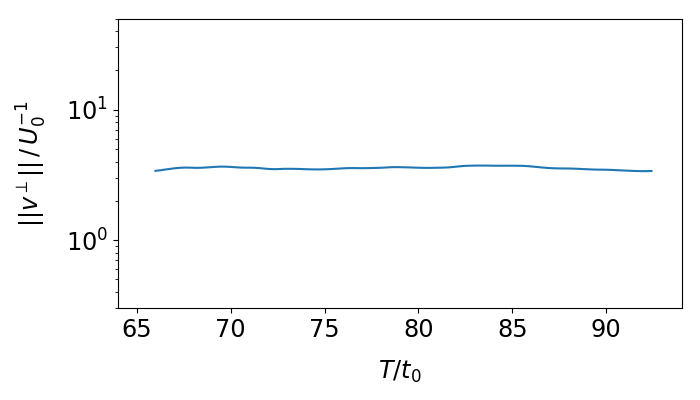}
    \caption{with $U$ as parameter.}
  \end{subfigure}
  \hfill
  \begin{subfigure}{0.49\textwidth}
    \includegraphics[trim=0cm 0cm 0cm 0cm, clip=true, width=\textwidth]{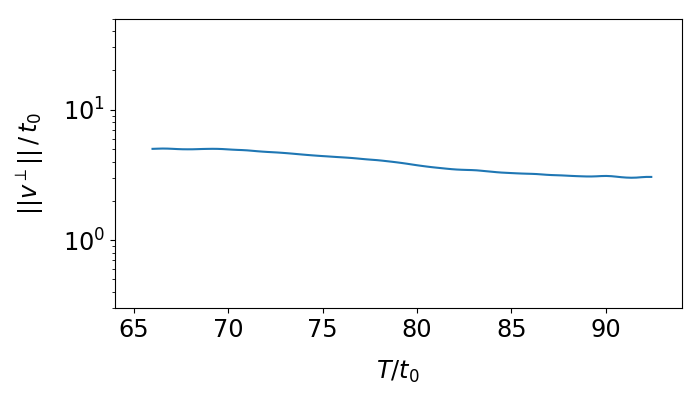}
    \caption{With $\omega$ as parameter.}
  \end{subfigure}
  \caption{Norms of shadowing directions on the finer mesh, measured from 66$t_0$ to 93$t_0$, computed on a trajectory of length $T=158t_0$.}
  \label{f:vperp norm}
\end{figure}

We first plot the norm of shadowing directions on the finer mesh in figure~\ref{f:vperp norm},
where the norm is induced by the inner product defined in equation~\eqref{e:define_inner_products}.
Since $v^\perp = du^\perp/ds$, with $du^\perp$ already normalized by free-stream constants in the definition of inner product,
we further need to normalize shadowing directions by $s_0^{-1}$. 
Here $s_0$ is the unit parameter, that is, $U_0$ or $\omega_0$.

By inspecting figure~\ref{f:vperp norm}, we can verify the uniform boundedness of shadowing directions,
as both $\|v^\perp\|$ do not change much from from $66t_0$ to $93t_0$.
For comparison we may think of the first CLV, whose norm would grow $e^{0.21\times(93-66)} = 290$ times larger within the same time span.

\begin{figure}
  \centering
  \begin{subfigure}{\textwidth}
    \includegraphics[trim=4cm 13cm 20cm 13cm, clip=true, width=0.49\textwidth]{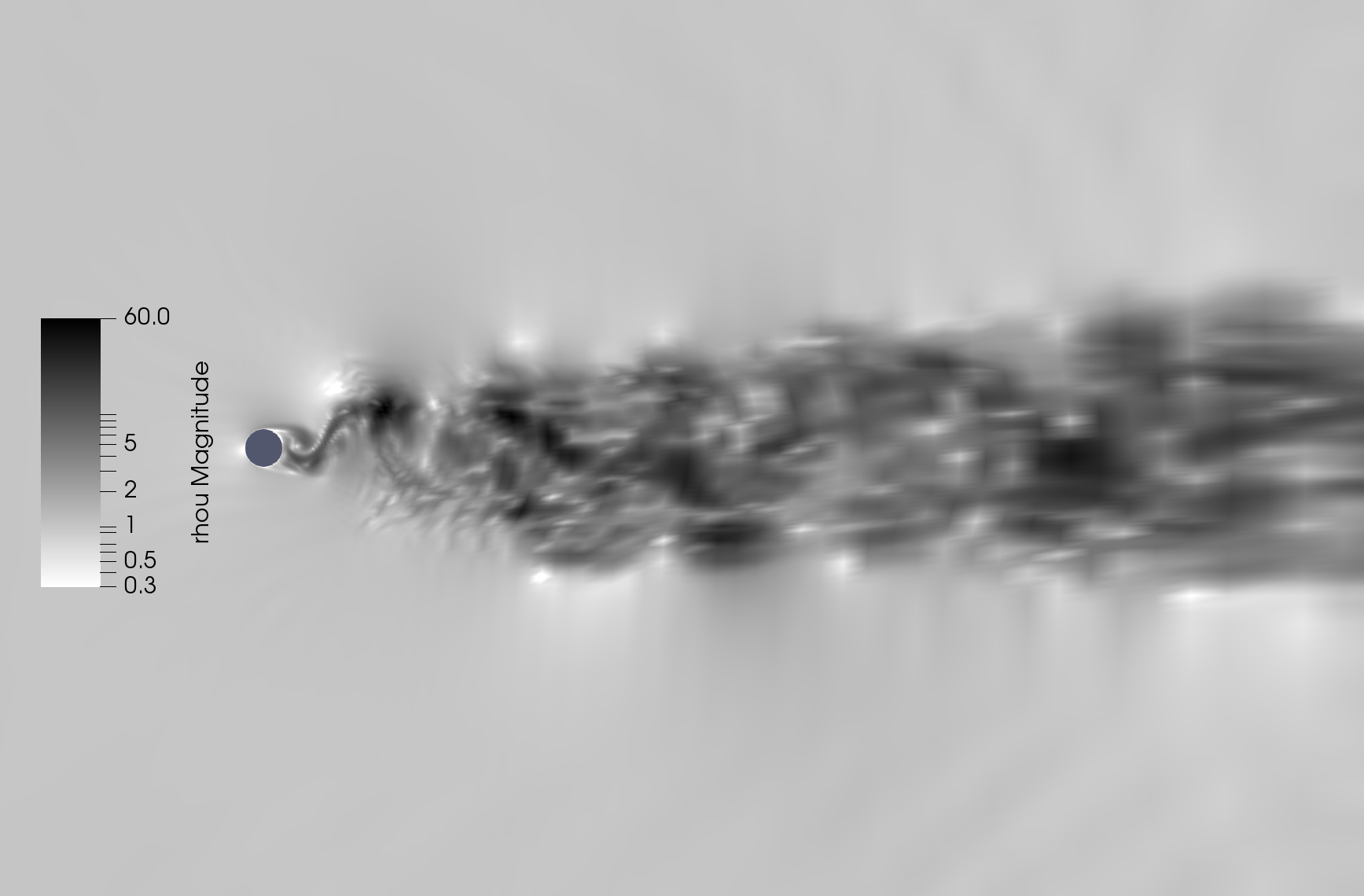}
    \includegraphics[trim=4cm 13cm 20cm 13cm, clip=true, width=0.49\textwidth]{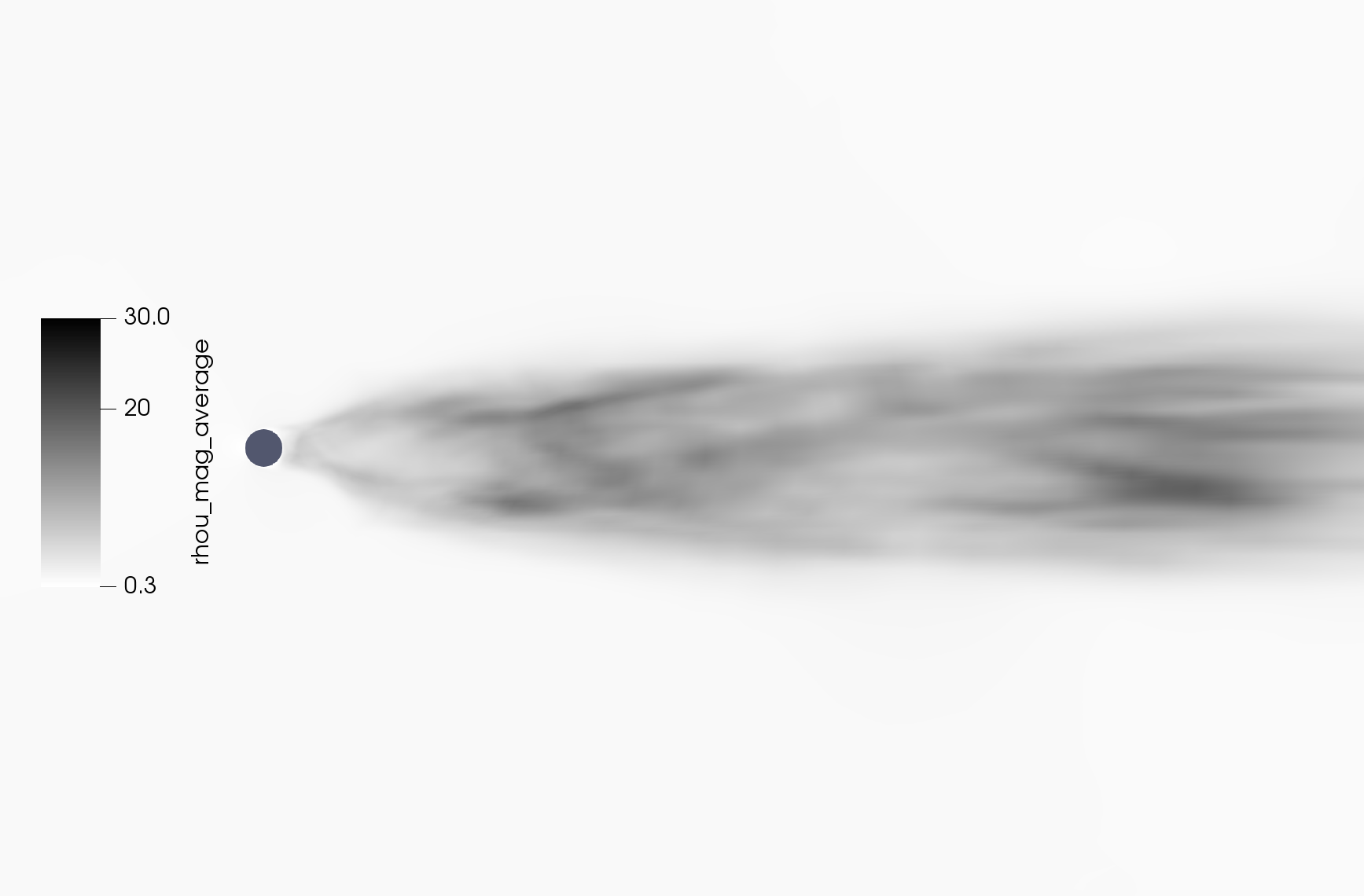}
    \caption{$U$ as the parameter, normalized by $\rho_0U_0/U_0=\rho_0$.}
    \label{f:u}
  \end{subfigure}

  \begin{subfigure}{\textwidth}
    \includegraphics[trim=4cm 13cm 20cm 13cm, clip=true, width=0.49\textwidth]{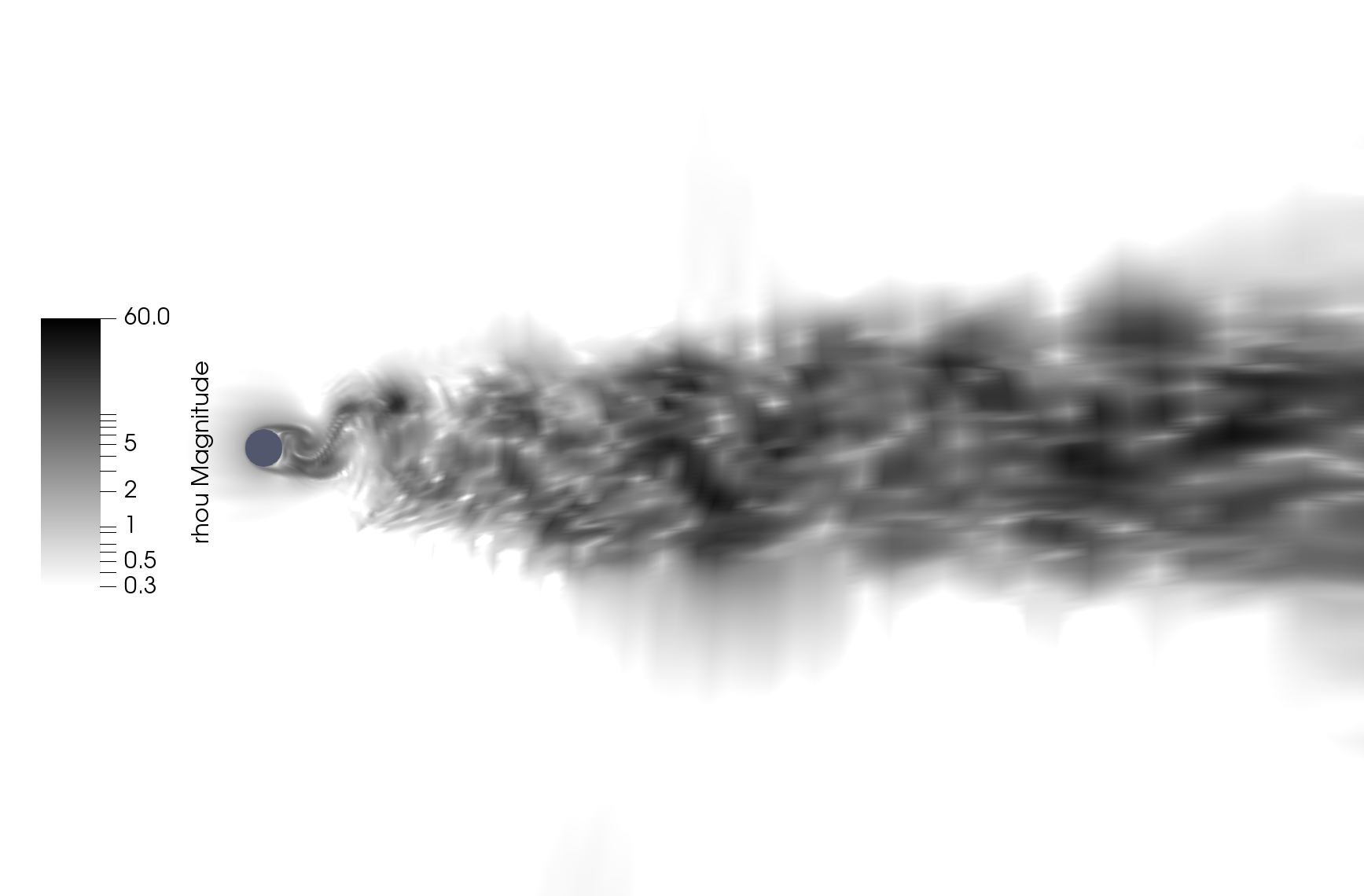}
    \includegraphics[trim=4cm 13cm 20cm 13cm, clip=true, width=0.49\textwidth]{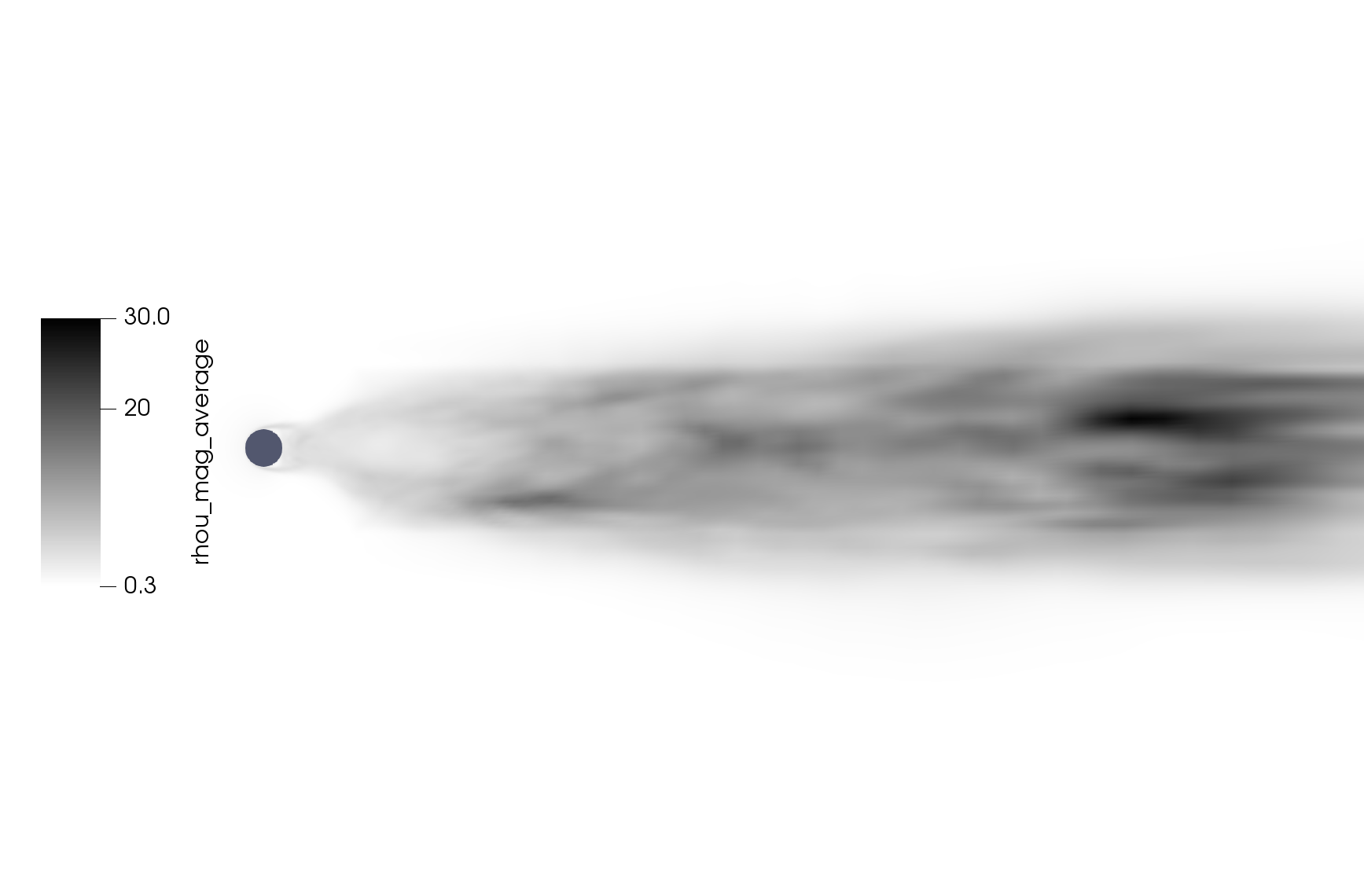}
    \caption{$\omega$ as the parameter, normalized by $\rho_0U_0/\omega_0$.}
    \label{f:w}
  \end{subfigure}

  \begin{subfigure}{\textwidth}
    \includegraphics[trim=4cm 13cm 20cm 13cm, clip=true, width=0.49\textwidth]{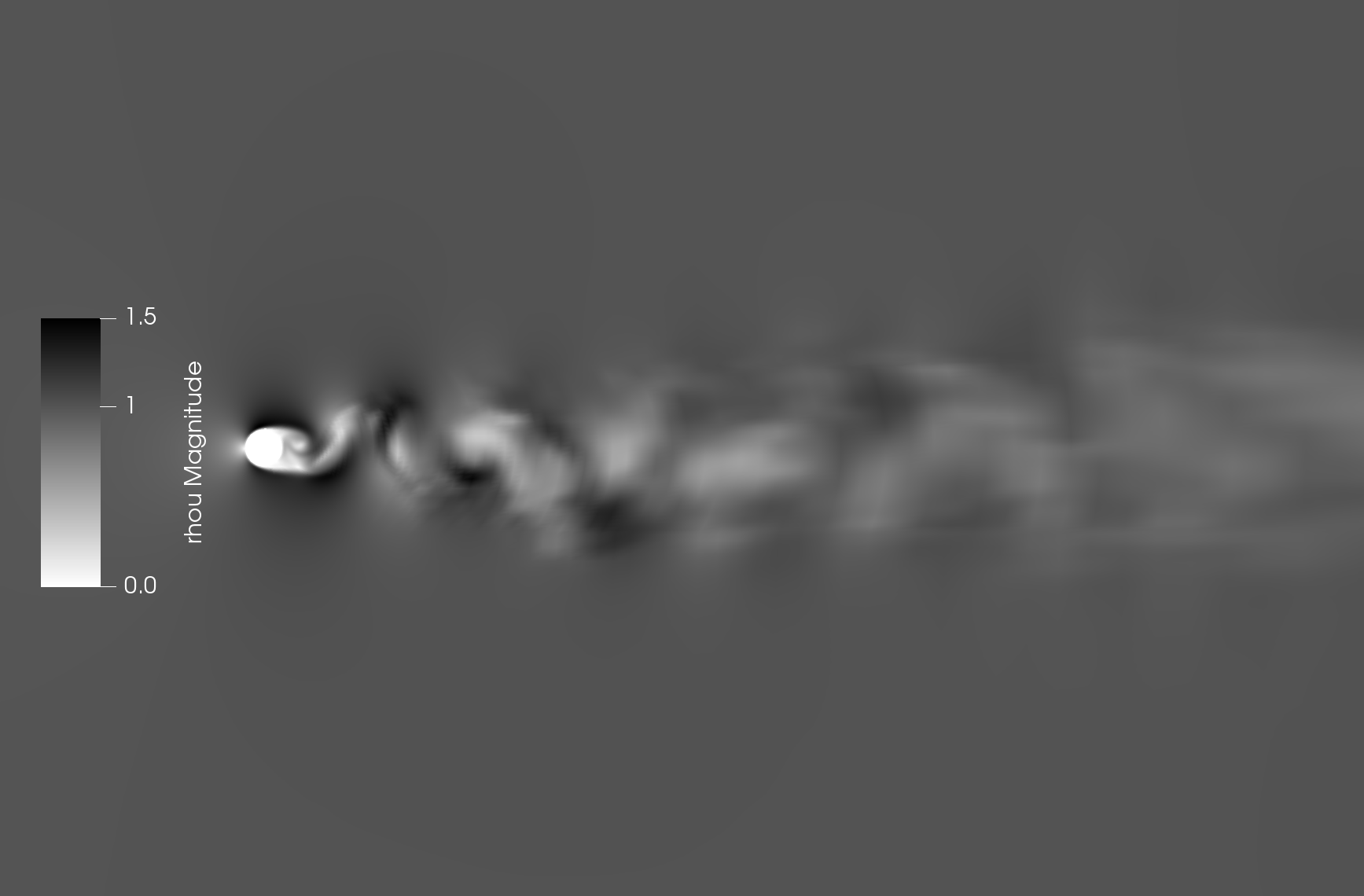}
    \includegraphics[trim=4cm 13cm 20cm 13cm, clip=true, width=0.49\textwidth]{figs/primal_averaged_finer.png}
    \caption{Primal flow field, normalized by $\rho_0U_0$.}
  \end{subfigure}
  
  \caption{Magnitude of the $\rho U$ component of the shadowing direction  $v^\perp = du^\perp/ds$ on the finer mesh, 
    which is computed by FD-NILSS on a trajectory of time length $T = 158t_0$.
  Left: a snapshot at $0.5T$, colored by log scale; right: averaged from $66t_0$ to $93t_0$, colored by linear scale.}
  \label{f:vperp field}
\end{figure}

Orthogonal projections of shadowing directions, $v^\perp$, are plotted in figure~\ref{f:vperp field}.
\footnote{Movies of $v^\perp$ on the finer mesh can be found on YouTube via 
\href{https://www.youtube.com/playlist?list=PLAlGrl2Jghfu31xrP3v_rnpiRRt3RHQnR}{this link}.}
Same as CLVs, for any time $t$, $v^\perp(t)$ lives in the same function space as the primal solution $u(t)$,
hence $v^\perp$ also has $\rho$, $\rho U$, and $\rho E$ components.
In figure~\ref{f:vperp field}, for each parameter, 
we plot the $|\rho U|$ field of $v^\perp(t)$.
Unlike CLVs, both the direction and the magnitude of shadowing directions are meaningful, 
since shadowing directions reflect not only in what direction flow fields will change due to perturbations in system parameters $s$,
but also how large the change will be.
Hence, we normalize shadowing directions by free-stream constants and unit parameters to preserve information about magnitudes.

If we perturb system parameters by $\Delta U$, the shadowing solution is a flow field with free-stream property $\rho_0 (U_0+\Delta U)$.
Since $v^\perp$ points from the base solution to the shadowing solution,
we expect the free-stream area of $v^\perp$ to be $\rho_0 \Delta U / \Delta U = \rho_0$.
The above intuition is verified on the left of figure~\ref{f:u}: with $U$ as the system parameter, 
the free-stream area of the $\rho U$ component of the shadowing direction $v^\perp$ has magnitude $1$, since we are normalizing by $\rho_0$.
On the other hand, perturbing rotation speed $\omega$ should have little effect on the free-stream area:
this is also verified in figure~\ref{f:w}, where $v^\perp$ has small magnitude in the free-stream area.

We observe by comparing the left of figure~\ref{f:u} and figure~\ref{f:w} that,
in the wake area, both $v^\perp$ have similar magnitude.
This is further confirmed by the fact that in figure~\ref{f:vperp norm} both $\|v^\perp\|$ are similarly large after normalization.
This fits our intuition that, if we perturb cylinder rotation by $\Delta \omega = 0.01\omega_0$,
then the circumferential speed is changed by $0.01\omega_0 D/2 = 0.01U_0/2$:
this extra rotation should have similar magnitude of impact on the wake as changing the free-stream velocity by $\Delta U = 0.01U_0$.

We compare the shadowing directions with the primal flow field in figure~\ref{f:vperp field}.
On one hand, shadowing directions are correlated to the primal flow field, 
since the peak values of shadowing directions are typically located close to the vortices,
indicating that vortices structures are the most sensitive to parameter perturbations.

On the other hand, because shadowing directions are perturbations, they are very different from the primal flow.
The primal flow has several structures that are insensitive to parameter perturbations,
whereas parameter perturbations may bring new structures to the primal flow.
For example, when $\omega$ is parameter, the stagnant area in front of the cylinder in primal flow 
changes to a smooth halo around the cylinder in the shadowing directions,
due to the fact that perturbing $\omega$ leads to a uniform acceleration around the cylinder.
Moreover, the shadowing direction has a larger active area than the primal flow.
An extreme example is the left of figure~\ref{f:u}, where the entire free stream of shadowing directions has value one, as we discussed.
Indeed, the convolution formula of shadowing directions shows that shadowing directions are affected by both stable and unstable CLVs 
\citep{Ni_adjoint_shadowing}, the union of whose active areas cover the entire flow field.

\subsection{Results of sensitivities}

With shadowing directions, 
we can compute sensitivities of long-time-averaged objectives 
from \eqref{e:xi end value} and \eqref{e:djds seg} with little extra cost.
In fact, as we shall see, using only a trajectory of time length 106$t_0$ should be enough to compute accurate sensitivities.
In this subsection, we investigate the effect of $U$ on the averaged drag force $\avg{D_r}$.
We will normalize $\avg{D_r}$ by $F_0=0.5\rho U_0^2 D Z=8.031\times 10^{-5}$.
The objective for system parameter $\omega$ is averaged lift $\avg{L}$, which is also normalized by $F_0$.

\begin{figure}
  \begin{subfigure}{0.49\textwidth}
    \includegraphics[trim=0cm 0cm 0cm 0cm, clip=true, width=\textwidth] {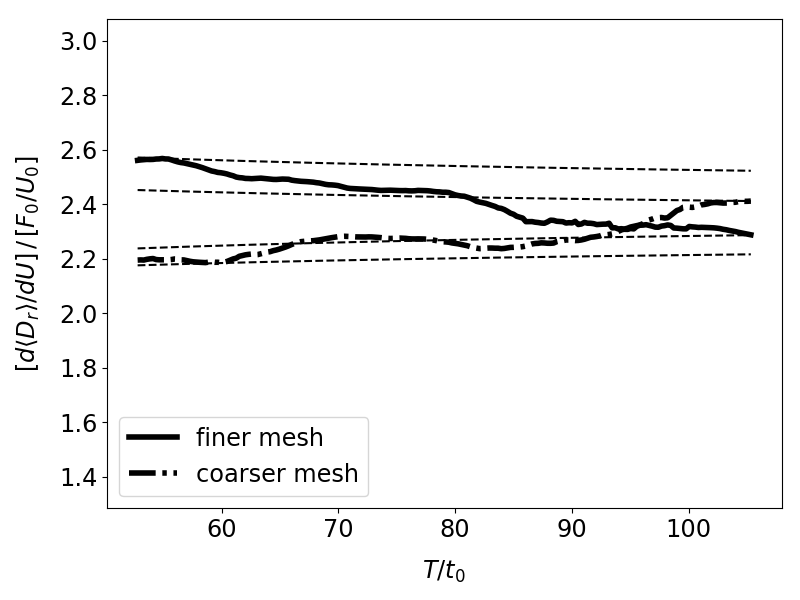}
    \caption{Drag versus $U$}
  \end{subfigure}
  \hfill
  \begin{subfigure}{0.49\textwidth}
    \includegraphics[trim=0cm 0cm 0cm 0cm, clip=true, width=\textwidth] {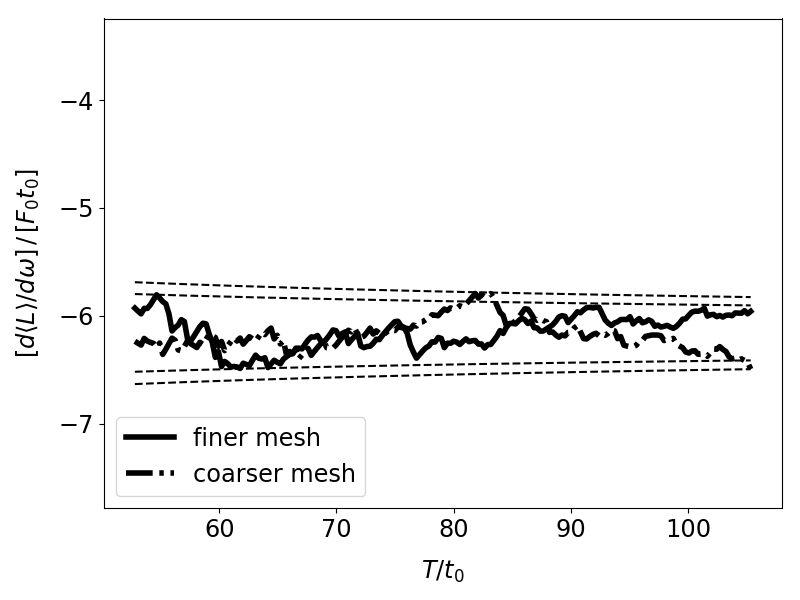}
    \caption{Lift versus $\omega$}
  \end{subfigure}
  \caption{History plots of sensitivities computed by FD-NILSS.}
  \label{f:djds history}
\end{figure}

We plot the history of the sensitivities computed via shadowing directions in figure~\ref{f:djds history}.
For each $T$, we run NILSS from the same initial condition of primal and tangent solutions to $T$, solve the shadowing direction,
using which we compute the sensitivity corresponding to $T$.
The uncertainty of the final sensitivity is estimated by the smallest interval 
which bounds the history of the sensitivity and whose size shrinks as $T^{-0.5}$.
We can see that for both meshes the uncertainty of sensitivities is below 10\%.

\begin{figure}
  \begin{subfigure}{0.49\textwidth}
    \includegraphics[trim=0cm 0cm 0cm 0cm, clip=true, width=\textwidth]{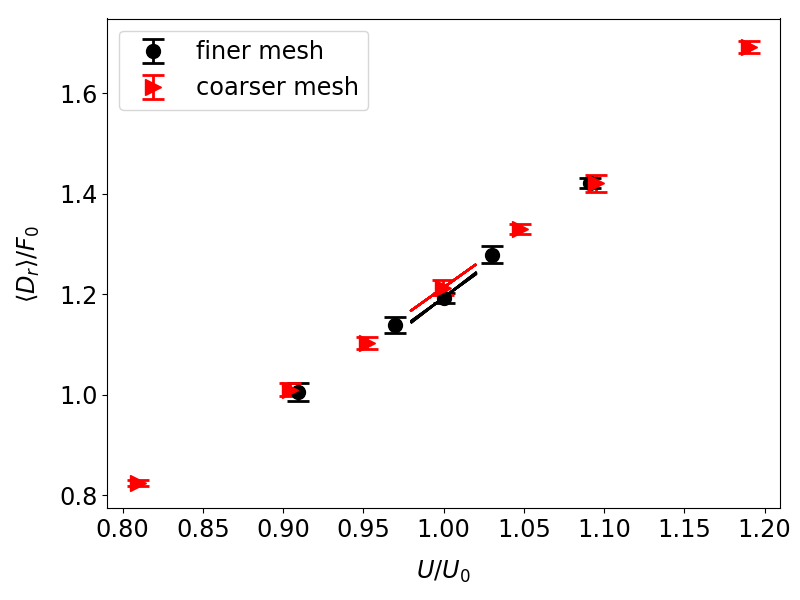}
    \caption{Drag versus $U$}
  \end{subfigure}
  \hfill
  \begin{subfigure}{0.49\textwidth}
    \includegraphics[trim=0cm 0cm 0cm 0cm, clip=true, width=\textwidth]{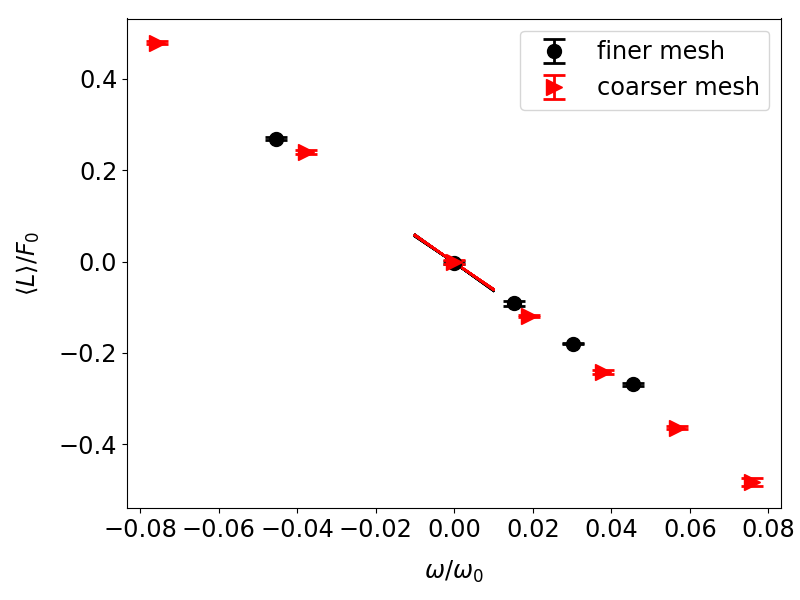}
    \caption{Lift versus $\omega$}
  \end{subfigure}

  \caption{Sensitivities for different choices of parameters and objectives. 
  Vertical bars indicate 95\% confidence intervals of the averaged objectives.
  The straight lines though the middle data points indicate the sensitivities computed by FD-NILSS.}
  \label{f:result djds}
\end{figure}

We explain how to compute the uncertainty of long-time-averaged objectives,
which is approximated by $\avg{J}_{T'}$, the averaged instantaneous objective over $T'=7.68\times 10 ^{-3} = 1014 t_0$.
To get the uncertainty due to taking the average over finite time, we divide the history of $J(t)$ into five equally long parts.
We denote the objectives averaged over each of the five parts by $J_1,...J_5$, whose corrected sample standard deviation is denoted by $\sigma'$.
We assume that the standard deviation of $\avg{J}_{T'}$ is proportional to $T'^{-0.5}$,
so we use $\sigma = \sigma' / \sqrt{5}$ as the standard deviation of $\avg{J}_{T'}$.
We further assume $\pm2\sigma$ yields the 95\% confidence interval, which is indicated by the vertical bar in figure~\ref{f:result djds}.

The sensitivities computed via shadowing directions are shown in figure~\ref{f:result djds}.
For both meshes our sensitivities correctly reflect the trend between averaged objectives and system parameters.
This confirms the physical meaning of the shadowing direction: 
it reveals the sensitivity of all physical phenomena in this flow field with respect to perturbations in parameters,
in such a way that the average of the sensitivity is the sensitivity of the average.

The cost of FD-NILSS for computing a sensitivity is mainly in integrating the primal solution over 
$400 \times 200 \times 32 = 2.6 \times 10 ^6$ time steps.
Here $400$ is the number of segments, $200$ is the number of time steps in each segment,
and $32$ is the number of primal solutions computed.
In FD-NILSS we need one inhomogeneous tangent and 30 homogeneous tangent solutions,
and each tangent solution is approximated by the difference between a perturbed solution and the same base solution:
that is 32 primal solutions in total.
On the other hand, a sensitivity can be revealed by some regression 
among the five pairs of objectives and parameters in figure \ref{f:result djds},
the total cost of which is $3.8\times 10 ^6$ steps of primal simulation: this cost is similar to that of FD-NILSS.

NILSS can be further accelerated when we have a tangent solver, 
by taking advantage of the fact that all adjoint solutions use the same Jacobian $\partial_u f$.
That is, we can integrate all tangent solutions simultaneously without repeatedly loading $\partial_u f$ into the computer CPU,
which is the most time-consuming procedure in the numerical integration.
In other words, we can perform, at each time step, one matrix-matrix product, where the second matrix is composed of several tangent solutions,
rather than performing several matrix-vector products, where we need to reload the matrix for each product.
\footnote{This idea was brought up during private discussions with Pablo Fernandez, who co-authored FD-NILSS.}

Additionally, as discussed in \citep{Ni_NILSS_JCP,Ni_fdNILSS}, FD-NILSS and NILSS have smaller marginal cost for new parameters;
for cases with more parameters than objectives or cases when we have only adjoint solvers, 
the non-intrusive least-squares adjoint shadowing (NILSAS) method \citep{Ni_nilsas} 
can compute the gradient of one objective with respect to many parameters in one run.

From the above discussions, together with the implication of the conjecture in section~\ref{s:CLV results}
that there should be a small fraction of unstable CLVs for open flows,
we believe that NILSS and NILSAS are competitive in terms of efficiency among all sensitivity algorithms.

As we can see, although our system is not uniform hyperbolic, 
the sensitivities computed by shadowing methods are still valid for both meshes.
We thus add one more piece of evidence to the chaotic hypothesis: that is,
although shadowing methods are logically consequential to uniform hyperbolicity, 
they are still valid for our fluid system, 
which only exhibits some hyperbolic phenomena (angles between far-apart CLVs are large), 
but is not uniform hyperbolic.

\section{Conclusions}

In this paper, we first compute the Lyapunov Exponents (LEs) of a 3D chaotic flow past a cylinder at Reynolds number 525.
Our computation shows that this flow problem has less than 30 positive LEs, but the detailed LE spectrum depends on meshes.
For both meshes, we find that the Lyapunov dimension is smaller than 109,
meaning that the apparently complicated dynamics of our 3D flow problem 
can be attributed to the interaction among less than 109 degrees of freedom.

We then compute the first 40 Covariant Lyapunov Vectors (CLVs) of this 3D flow problem.
We find that the angles between CLVs become larger when the indices are further apart, 
although there may be occasional tangencies between adjacent CLVs.
Our conjecture is backed by plots of CLVs, 
where we find that unstable CLVs are active in the instability-generating areas such as boundary layers and near wakes,
while stable CLVs are active in more dissipative areas such as the far wake, the front of the cylinder, and the free stream.
This difference in active areas is robust to meshes, and it indicates that CLVs point to different directions, 
and it also implies that for open flows there is a large fraction of CLVs that are stable.

We also conclude that our system is not uniform hyperbolic.
Firstly, due to the extra neutral CLV corresponding to translation along the spanwise direction, our problem has at least two neutral directions.
Hence at least the first assumption in uniform hyperbolicity is violated on both meshes.
Secondly, we found that the smallest angle between CLVs is not robust to meshes and time length, 
since on the finer mesh the smallest angle is well above the threshold value, 
whereas on the coarser mesh it almost equals the threshold.

We then use the finite difference non-intrusive least-squares shadowing (FD-NILSS) algorithm 
to compute the shadowing direction $v^{\perp}$ of this 3D flow problem.
The norm of the computed $v^{\perp}$ remains on the same level as time evolves, 
thus indicating the existence of shadowing directions of this 3D flow problem.
By observing normalized $v^{\perp}$, 
we find that its value in the free-stream area is roughly one for the perturbation on the free-stream speed $U$,
but zero for the perturbation on the cylinder rotation speed $\omega$.
On the other hand, normalized $v^\perp$ have similar magnitudes in the wake area for perturbations on $U$ and $\omega$.

Finally, we use shadowing directions to compute the sensitivities of some long-time-averaged objectives.
For both meshes, the computed sensitivities correctly reflect the trend between system parameters and objectives.
Our results confirm the physical meaning of the shadowing solution in revealing 
sensitivities of all physical phenomena in this flow field with respect to perturbations on a parameter.
The sensitivity results also suggest that shadowing methods are robust to numerical implementations,
and may be valid for more general chaotic fluid systems, although shadowing theories are mathematically developed under uniform hyperbolicity.

\bibliographystyle{jfm}
\bibliography{MyCollection}
\end{document}